\documentclass[12pt,a4paper]{article}
\pdfoutput=1

\usepackage[utf8]{inputenc}
\usepackage[british]{babel}
\usepackage{authblk}
\usepackage{amsmath,amssymb}
\usepackage{bbold}
\usepackage{booktabs}
\usepackage{tabularx}
\usepackage{graphicx}
\usepackage{cite}
\usepackage{subcaption}

\usepackage{hyperref}
\usepackage{hypcap}

\allowdisplaybreaks
\usepackage[a4paper,
margin=2cm
]{geometry}

\newcommand{\myparagraph}[1]{\paragraph{#1}\mbox{}\\}
\def\A{k}
\def\B{l}
\def\C{m}
\def\D{n}
\def\ts{$\times$}

\def\elltau{\text{$\ell$--$\tau$}}
\def\tauh{\tau h}

\newcommand{\MR}{M_\text{S}}
\newcommand{\MS}{M_\text{S}}

\hypersetup{%
colorlinks,citecolor=blue,urlcolor=blue,linkcolor=blue,%
hypertexnames=false,%
pdftitle={Constraints on leptoquarks from lepton-flavour-violating tau-lepton processes},%
pdfauthor={T. Husek, K. Monsalvez-Pozo, J. Portoles}%
}

\title{\bf\boldmath Constraints on leptoquarks from lepton-flavour-violating tau-lepton processes}
\author[a,b]{\sc\small Tom\'{a}\v{s} Husek\thanks{Email:~tomas.husek@thep.lu.se}}
\author[a]{\sc\small Kevin Mons\'alvez-Pozo\thanks{Email:~kevin.monsalvez@ific.uv.es}}
\author[a]{\sc\small Jorge~Portol\'es\thanks{Email:~jorge.portoles@ific.uv.es}}
\affil[a]{\footnotesize Instituto de F\'isica Corpuscular, CSIC -- Universitat de Val\`encia,}
\affil[{ }]{Apt.\ Correus 22085, E-46071 Val\`encia, Spain}
\affil[{ }]{\tiny}
\affil[b]{\footnotesize Department of Astronomy and Theoretical Physics, Lund University,}
\affil[{ }]{Box 43, SE-22100 Lund, Sweden}
\date{}

\begin{document}

\ \vskip-1.5cm\hfill {\small LU TP 21-48}\\
{\let\newpage\relax\maketitle}

\abstract{
Leptoquarks are ubiquitous in several extensions of the Standard Model and seem to be able to accommodate the universality-violation-driven $B$-meson-decay anomalies and the $(g-2)_\mu$ discrepancy interpreted as deviations from the Standard Model predictions. In addition, the search for lepton-flavour violation in the charged sector is, at present, a major research program that could also be facilitated by the dynamics generated by leptoquarks. In this article, we consider a rather wide framework of both scalar and vector leptoquarks as the generators of lepton-flavour violation in processes involving the tau lepton. We single out its couplings to leptoquarks, thus breaking universality in the lepton sector, and we integrate out leptoquarks at tree level, generating the corresponding dimension-6 operators of the Standard Model Effective Field Theory. In Ref.~\cite{Husek:2020fru} we obtained model-independent bounds on the Wilson coefficients of those operators contributing to lepton-flavour-violating hadron tau decays and $\ell$--$\tau$ conversion in nuclei, with $\ell=e,\mu$. Hence, we use those results to translate the bounds into the couplings of leptoquarks to the Standard Model fermions.
}

\section{Introduction}
\label{sec:intro}

The Standard Model (SM) of particle physics is a very successful quantum field theory, which describes the dynamics of the strong interaction as well as the unified electromagnetic and weak interactions --- the electroweak (EW) theory.
While the SM has passed a number of elaborate experimental tests over a broad range of energies, it is believed already for decades that it does not provide us with the final and complete picture of reality as expected from a fundamental theory.
There are purely theoretical reasons to think like that: Besides the fact that the SM contains many a priori unknown parameters, there are indications for the unification of the strong and EW forces and the common underlying structure of all fermions, which form the so-called families or generations.
From the phenomenological point of view, there are several phenomena which cannot be explained within its framework.
For instance, the SM does not provide a viable dark-matter candidate, fails to predict the observed matter--antimatter asymmetry in the Universe, and, as is, it does not strive to unambiguously incorporate the tiny (though nonzero) masses of neutrinos.
Hence, one of the major goals of contemporary particle physics is to look beyond the SM (BSM) for possible explanations of these and other shortcomings.
\par
The effects of BSM phenomena on the dynamics of SM particles, arising at energy scales higher than the EW scale ($\Lambda_\text{EW}$), can be encoded in terms of the Standard Model Effective Field Theory (SMEFT)~\cite{Buchmuller:1985jz,Grzadkowski:2010es}, in particular in the Wilson coefficients (WCs) --- low-energy constants standing in front of the monomials in such an effective Lagrangian.
These coefficients can be related to parameters of particular BSM models and could be also determined from experimental results.
\par 
Within the rich palette of BSM scenarios, a very well motivated class of theories predicts the existence of leptoquarks (LQs) --- electrically charged bosons (with spin $S=0,1$) which transform as triplets under SU(3)$_\text{C}$ and can turn quarks into leptons and vice versa.
They naturally emerge in Grand Unified Theories, where strongly non-interacting leptons are accommodated into the same multiplets as quarks: They first appeared in the Pati--Salam model~\cite{Pati:1973uk,Pati:1974yy}, and right after in theories based on other symmetry groups, such as the simplest SU(5) in the case of the Georgi--Glashow model~\cite{Georgi:1974sy}, SO(10)~\cite{Georgi:1974my,Fritzsch:1974nn}, or further on in superstring-inspired E$_6$ models~\cite{Witten:1985xc,Hewett:1988xc}.
They were as well predicted in technicolor and other related models based on the dynamically generated symmetry breaking~\cite{Dimopoulos:1979es,Ellis:1980hz,Farhi:1980xs}, or in models with composite fermions~\cite{Schrempp:1984nj,Buchmuller:1985nn,Gripaios:2009dq} or extended scalar sectors~\cite{Davies:1990sc,Arnold:2013cva}.
\par 
At the same time, the persisting existence of several anomalies --- discrepancies between the SM theoretical predictions of observables and their experimental values --- signals possible effects of new physics (NP), and leptoquarks present themselves as relevant NP candidates, being able to address one or more of these deviations, depending on the chosen model.
The discrepancies that have drawn more attention in the recent literature are $R_{D^{(*)}}$~\cite{Matyja:2007kt,Lees:2012xj,Lees:2013uzd,Huschle:2015rga,Aaij:2015yra,Aaij:2017deq,Aaij:2017uff,Abdesselam:2019dgh}, $R_{K}$~\cite{Aaij:2014ora,CMS:2014xfa,Aaij:2015oid,Abdesselam:2016llu,Aaij:2017vbb,Aaij:2019wad,Aaij:2020nrf,Perez:2021ddi} together with very recent $R_{K_{S}^{0}}$ and $R_{K^{*+}}$~\cite{LHCb:2021lvy}, the so-called $B$ anomalies~\cite{London:2021lfn}, and the anomalous magnetic moment ($g-2$) of the muon~\cite{Bennett:2006fi,Abi:2021gix}.
Effects of LQs in these and other processes are extensively studied and parameterised via effective field theory (EFT) frameworks; see e.g.\ Ref.~\cite{Dorsner:2016wpm} and references therein.
More recent updates on the role of leptoquarks in $B$ anomalies and constraints on their couplings to ordinary matter can be found in Refs.~\cite{Popov:2016fzr,Cheung:2017efc,Crivellin:2017zlb,Cai:2017wry,Buttazzo:2017ixm,Sahoo:2018ffv,Becirevic:2018afm,Crivellin:2018yvo,Biswas:2018snp,Angelescu:2018tyl,Bansal:2018nwp,Iguro:2018vqb,Aebischer:2018acj,Cornella:2019hct,Yan:2019hpm,Popov:2019tyc,Crivellin:2019szf,Crivellin:2019dwb,Saad:2020ihm,Gherardi:2020qhc,Babu:2020hun,Bordone:2020lnb,Crivellin:2020mjs}, while the $(g-2)_{\mu}$ discrepancy is addressed in Refs.~\cite{Davidson:1993qk,Babu:2010vp,Popov:2016fzr,Cheung:2017efc,Crivellin:2017zlb,Cai:2017wry,Buttazzo:2017ixm,Crivellin:2018qmi,Kowalska:2018ulj,Mandal:2019gff,Dorsner:2019itg,Crivellin:2019dwb,Bigaran:2020jil,Bigaran:2021kmn,Davier:2019can,Saad:2020ihm,DelleRose:2020qak,Gherardi:2020qhc,Bordone:2020lnb,Crivellin:2020mjs,Zhang:2021dgl}.
A detailed analysis of low-energy signals of scalar leptoquarks is presented in Ref.~\cite{Mandal:2019gff}.
Finally, for current and expected limits from collider searches see Refs.~\cite{Bandyopadhyay:2018syt,Hiller:2018wbv,Monteux:2018ufc,Becirevic:2018afm,Crivellin:2020ukd,Biswas:2018snp,Angelescu:2018tyl,Schmaltz:2018nls,Cerri:2018ypt,Faber:2018afz,Cornella:2019hct,Zhang:2019jwp,Chandak:2019iwj,Allanach:2019zfr,Borschensky:2020hot,Buonocore:2020erb,ATLAS:2020dsk,Crivellin:2020mjs,Bhaskar:2021gsy}.
\par 
Besides the notorious anomaly-related issues, leptoquarks have also been considered to address other BSM problems like the generation of neutrino masses through one~\cite{Mahanta:1999xd,Deppisch:2016qqd,Popov:2016fzr,Popov:2019tyc,Bigaran:2019bqv,Zhang:2021dgl}, two~\cite{Babu:2010vp,Popov:2016fzr,Cai:2017wry,Saad:2020ihm,BhupalDev:2020zcy} and three loops~\cite{Cheung:2017efc}.
Furthermore, their role as mediators between the SM sector and dark matter candidates is studied in Refs.~\cite{Mandal:2018czf,Mohamadnejad:2019wqb}, their implications for baryogenesis are considered in Ref.~\cite{Hati:2018cqp}, and the ANITA anomalous events are explained using particular leptoquark models in Refs.~\cite{Chauhan:2018lnq,BhupalDev:2020zcy}.
Their existence might also offer a hint on why there are exactly three generations of matter or why there is the same number of quark and lepton species, the consequence of which is the fact that the currents associated with the SM gauge symmetries are non-anomalous --- free of Adler--Bell--Jackiw axial anomalies~\cite{Adler:1969gk,Bell:1969ts,Bouchiat:1972iq}.
\par 
Following the above reasoning, leptoquarks belong, at present, among the most promising NP contributions.
However, despite the immense experimental effort, they have not been directly observed yet.
\par
In this work, we address another interesting BSM phenomenon, namely processes exhibiting charged-lepton-flavour violation (CLFV).
This effect, while absent in the SM, is expected to happen in presence of massive neutrinos.
However, minimal extensions of the SM with light right-handed neutrinos predict tiny CLFV rates, inaccessible to current and mid-term foreseen experiments~\cite{Cheng:1977nv,Petcov:1976ff,Bilenky:1977du,Marciano:1977wx,Lee:1977qz,Lee:1977tib}: e.g.\ $\Gamma(\mu \rightarrow e \gamma) / \Gamma(\mu \rightarrow e \nu \bar{\nu}) < 10^{-40}$.
Within the leptoquark framework, CLFV processes can occur at tree level via the exchange of a LQ coupled to $(\bar{\ell} \,\Gamma \,q)$ and
$(\bar{q} \,\Gamma \, \ell')$ currents (here, $q$ is short for a quark field, $\ell$ for a lepton field, and $\Gamma$ the relevant Dirac tensors), providing enhanced rates for these processes that could be measured at present or future experiments.
We will focus on CLFV $\tau$-involved processes since most of the work done in this area has been mainly related to the first and second families (see, for instance, the reviews~\cite{Raidal:2008jk,deGouvea:2013zba,Calibbi:2017uvl}), and the persistence of some charged-current-driven $B$ anomalies suggests an apparent violation of universality around the third family.
\par 
Hence, we take the most general couplings of the 5 different types of both scalar and vector leptoquarks to the SM fermions (see, for instance, Ref.~\cite{Dorsner:2016wpm}); in the presence of a right-handed neutrino, that we do not consider, there is an additional type of scalar and vector LQs.
Upon integration of those leptoquarks --- assuming $M_\text{LQ} \gg \Lambda_\text{EW}$ --- the four-fermion (mass-)dimension-6 ($D=6$) operators of the SMEFT \cite{Buchmuller:1985jz,Grzadkowski:2010es} are generated.
As commented above, we break universality in the lepton sector by attaching different couplings to the tau lepton and those of the first two families.
This is because some works~\cite{Bernigaud:2021fwn,Allwicher:2022vbf} point to the aforementioned particularity of the third family and, moreover, the implications related to the first two families in this regard are far less compelling~\cite{CMS:2021ctt}.
Moreover, we have also taken into account the $\tau \rightarrow \ell \gamma$ decays (with $\ell=e,\mu$), although their leading contribution arises at one loop.
Throughout this procedure, we get the WCs of the $D=6$ SMEFT operators expressed in terms of products of a pair of unknown couplings of LQs to SM fermions.
In addition, we identify the energy scale of the corresponding SMEFT with the masses of leptoquarks.
\par 
In Ref.~\cite{Husek:2020fru}, we performed a model-independent analysis taking into account current and foreseen experimental data for lepton-favour-violating hadronic $\tau$ decays (from Belle~\cite{HFLAV:2019otj} and Belle II~\cite{Kou:2018nap}) and $\ell$--$\tau$ conversion in nuclei, $\ell=e,\mu$ (from NA64 expected sensitivity~\cite{Gninenko:2018num}).
As a consequence, we obtained tight bounds on the participating WCs that we can now translate to the products of LQs couplings.
Upon general assumptions on those couplings we can also arrive at estimates for the lower bounds on LQ masses.
Based on our results, we would like to highlight the strong bounds on LQ masses and couplings that Belle II future results on the hadronic $\tau$ decays will be able to establish.
\par 
The paper is organised in the following way.
In Section~\ref{sec:LQLagrangian}, we present the most general CLFV leptoquark Lagrangian based on the SM symmetries accommodating scalars and vectors, and describe the important features of this framework.
In Section~\ref{sec:EFT}, we recover the four-fermion $D=6$ SMEFT operators that result from integrating out the LQ fields at tree level.
Hence, we give the relations between the WCs and the leptoquark couplings, which we use to constrain the latter in Section~\ref{sec:Results} by using our results from Ref.~\cite{Husek:2020fru}.
We point out our main conclusions in Section~\ref{sec:Conclusions}.
Several technical appendices make easier the understanding of the present work.

\section{Leptoquark Lagrangian}
\label{sec:LQLagrangian}

To systematically explore all possible options, leptoquarks are classified with respect to their spin (scalar or vector) and the way they couple to quarks and leptons based on their transformation properties under the SM gauge group $\text{SU}(3)_\text{C} \times \text{SU}(2)_\text{L} \times \text{U}(1)_\text{Y}$:
They always transform as colour triplets and range from SU(2) singlets to triplets; leptoquarks with the same SU(2) gauge dimensionality then differ by hypercharge.
The electric charge of the leptoquarks is then given, as usual, by $Q = I_3 + Y$, where $I_3$ stands for the SU(2) generator and $Y$ for the U(1) hypercharge operator.
Leptoquarks have a well-defined fermion number $F = 3 \, B + L$, with $B$ and $L$ being the baryon and lepton numbers, respectively.
All leptoquark fields are categorised into two sets: $|F| = 0,2$.
In what follows, we use the same notation and conventions as in Ref.~\cite{Dorsner:2016wpm}.
\par 
Generically, the kinetic and gauge couplings of leptoquarks are described by the Lagrangians
\begin{equation}
\label{eq:lsv}
\begin{split}
{\cal L}_\text{S} &= \sum_\text{scalars} \left[\, (D_{\mu} S)^{\dagger} \, D^{\mu} S \,  - \, M_S^2 \, S^{\dagger} S \, \right], \\
{\cal L}_\text{V} &= \sum_\text{vectors} \left[\,  - \frac{1}{2} \, V_{\mu \nu}^{\dagger} \, V^{\mu \nu} \,  + \, M_V^2 \, V^{\dagger}_{\mu}  V^{\mu}  \, +  \, \dots \, \right],
\end{split}
\end{equation}
where the field-strength tensor for the vector leptoquarks is $V_{\mu \nu} = D_{\mu} V_{\nu} - D_{\nu} V_{\mu}$.
The SM covariant derivative is given by
\begin{equation}
\label{eq:CovDer}
D_{\mu}=\partial_{\mu}+ig_1YB_{\mu}+ig_2I_{k}W_{\mu}^{k}+ig_3\frac{\lambda^{A}}{2}G_{\mu}^{A}\,,
\end{equation}
where the $\lambda^{A}$ and $I_{k}$ are the generators of the SU(3) and SU(2) symmetry groups, respectively, while $Y$ is the LQ hypercharge operator.
Note that the $I^{k}$ depend on the leptoquark SU(2) representation, e.g.\ for a leptoquark doublet, $I_{k}=\tau_{k}/2$, with $\tau_{k}$ being the Pauli matrices, while for a triplet we have $I_k = (I_k)_{lm} = - i \varepsilon_{klm}$, with $\varepsilon_{abc}$ being the three-dimensional Levi-Civita pseudotensor ($\varepsilon_{123}= 1$).
In Eq.~(\ref{eq:lsv}), the dots in the vector leptoquark Lagrangian correspond to other $D=4$ terms that involve additional interactions of the SM gauge fields with the leptoquarks, and self-interactions between the latter.
These are allowed by the gauge symmetry (although their couplings are not determined by it) and facilitate the renormalizability of the vector Lagrangian (see Ref.\cite{Gabrielli:2015hua}), but their explicit form is not essential for further discussion so we do not show them here.
Note that we do not consider these vector leptoquarks being gauge bosons of an extended gauge symmetry:
The gauge interactions of vector leptoquarks cannot be unambiguously defined due to their uncertain gauge nature, and an ultraviolet completion might be needed \cite{Gonderinger:2010yn}.
\par 
Finally, we do not consider the coupling of leptoquarks to the SM Higgs doublet \cite{Hirsch:1996qy, Crivellin:2020mjs}.
The diagonal couplings contribute, together with the mass parameters in the Lagrangian, to the total LQ mass after the spontaneous symmetry breaking has taken place and the Higgs acquired a vacuum expectation value.
On the other hand, the off-diagonal couplings translate into a mixing of leptoquarks in Table~\ref{tab:LQ} when diagonalising the total LQ mass matrix and working in the mass basis.
However, and as a consequence of the fact that the LQ--Higgs couplings involve two leptoquark fields, the corresponding mixing components behave as ${\cal O}(1/M_\text{LQ}^{2})$~\cite{Crivellin:2020mjs}.
Hence, upon integration, the effects of the induced mixing to the four-fermion operators present in our analysis are ${\cal O}(1/M_\text{LQ}^{4})$, and are thus of higher order than the ones we consider in our work.
\par 
The most general renormalizable Lagrangian based on the SM symmetries that realises interactions of leptoquarks with fermion pairs contains in total 10 types of leptoquark fields (which extends to 12 if right-handed neutrinos are brought into the picture): 5 scalar and 5 vector ones.
For each leptoquark type, the terms potentially responsible for the $\ell$--$\tau$ conversion and $\tau \rightarrow(\ell+\text{hadrons})$ decays (with $\ell=e,\mu$) are shown in Table~\ref{tab:LQ}, where all possible flavour structures for the Yukawa-like couplings should be taken into account.
\begin{table}[!t]
\centering
\renewcommand{\arraystretch}{1.5}
\begin{tabular}[th]{c|c|c}
\toprule
LQ type & SM symmetries & Lagrangian\\
\midrule
$S_3$ & $(\mathbf{\bar{3},3},1/3)$ & $Y^\text{LL}_{3,ij} \, \bar{Q}_\text{L}^{\text{C} i,a} \, \varepsilon^{ab} \, (\tau_k S_3^{k})^{bc} \, L_\text{L}^{j,c} + \text{h.c.}$ \\
$R_2$ & $(\mathbf{3,2},7/6)$ & $-Y^\text{RL}_{2,ij} \,\bar{u}_\text{R}^{i} \, R_2^{a} \, \varepsilon^{ab} \, L_\text{L}^{j,b} \, + \, Y^\text{LR}_{2,ij} \, \bar{e}_\text{R}^{i} \, R_2^{a \dagger} \, Q_\text{L}^{j,a} + \text{h.c.}$ \\
$\tilde{R}_2$ & $(\mathbf{3,2},1/6)$ & $-\tilde{Y}^\text{RL}_{2,ij} \, \bar{d}_\text{R}^{i} \, \tilde{R}_2^{a} \, \varepsilon^{ab} \, L_\text{L}^{j,b} + \text{h.c.}$ \\
$\tilde{S}_1$ & $(\mathbf{\bar{3},1},4/3)$ & $\tilde{Y}^\text{RR}_{1,ij} \, \bar{d}_\text{R}^{\text{C} ,i} \, \tilde{S}_1 \, e_\text{R}^{j} + \text{h.c.}$ \\
$S_1$ & $(\mathbf{\bar{3},1},1/3)$ & $Y^\text{LL}_{1,ij} \, \bar{Q}_\text{L}^{\text{C} i,a} \, S_1 \, \varepsilon^{ab} \, L_\text{L}^{j,b} \, + \, Y^\text{RR}_{1,ij}\, \bar{u}_\text{R}^{\text{C} i} \, S_1 \, e_\text{R}^{j} + \text{h.c.}$ \\
\midrule
$U_3$ & $(\mathbf{3,3},2/3)$ & $X^\text{LL}_{3,ij} \, \bar{Q}_\text{L}^{i,a} \, \gamma^{\mu} \, (\tau_k U_{3,\mu}^{k})^{ab} \, L_\text{L}^{j,b} + \text{h.c.}$ \\
$V_2$ & $(\mathbf{\bar{3},2},5/6)$ & $X^\text{RL}_{2,ij} \, \bar{d} \, _\text{R}^{\text{C} i}  \, \gamma^{\mu}  \, V_{2, \mu}^{a}  \, \varepsilon^{ab}  \, L_\text{L}^{j,b}  \, +   \, X^\text{LR}_{2,ij}  \, \bar{Q}_\text{L}^{\text{C} i, a}  \, \gamma^{\mu}  \, \varepsilon^{ab} \, V_{2,\mu}^{b}  \, e_\text{R}^{j} + \text{h.c.}$ \\
$\tilde{V}_2$ & $(\mathbf{\bar{3},2},-1/6)$ & $ - \,\tilde{X}^\text{RL}_{2,ij}  \, \bar{u}_\text{R}^{\text{C},i}  \, \gamma^{\mu}  \, \tilde{V}_{2, \mu}^{a} \, \varepsilon^{ab}  \, L_\text{L}^{j,b} + \text{h.c.}$ \\
$\tilde{U}_1$ & $(\mathbf{3,1},5/3)$ & $\tilde{X}^\text{RR}_{1,ij}  \, \bar{u}_\text{R}^{i}  \, \gamma^{\mu}  \, \tilde{U}_{1, \mu}  \, e_\text{R}^{j} + \text{h.c.}$ \\
$U_1$ & $(\mathbf{3,1},2/3)$ & $X^\text{LL}_{1,ij}  \, \bar{Q}_\text{L}^{i, a}  \, \gamma^{\mu}  \, U_{1, \mu}  \, L_\text{L}^{j, a}  \, +  \, X^\text{RR}_{1,ij} \, \bar{d}_\text{R}^{i}  \, \gamma^{\mu}  \, U_{1, \mu}  \, e_\text{R}^{j} + \text{h.c.}$\\
\bottomrule
\end{tabular}
\caption{
\label{tab:LQ}
Classification of leptoquarks based on the representations of matter fields under the SM gauge group $\text{SU}(3)_\text{C} \times \text{SU}(2)_\text{L} \times \text{U}(1)_\text{Y}$, and related Lagrangians representing the interactions of leptoquarks with SM quarks and leptons.
Only terms potentially responsible for CLFV are shown, and all these terms then constitute the ${\cal L}_\text{LQ--SM}$ Lagrangian.
The Yukawa-like couplings $Y_{d,ij}^{\chi_1 \chi_2}$ and $X_{d,ij}^{\chi_1 \chi_2}$, $d=1,2,3$, are dimensionless.
The hypercharge $Y$ is related to the electric charge $Q$ via $Q = I_3 + Y$, with $I_3$ staying for the third SU(2)$_\text{L}$ generator, the specific realisation of which, as mentioned in the main text, depends on the corresponding leptoquark representation.
As is customary, $Q_\text{L}$ and $L_\text{L}$ stand for the left-handed quark and lepton SU(2) doublets, $u_\text{R}(d_\text{R})$ and $e_\text{R}$ are the up(down)-type quark and lepton right-handed SU(2) singlets, respectively.
Finally, $\tau_k$ are the Pauli matrices ($\{\tau_k,\tau_l\}=2\delta_{kl}\mathbb{1}_2$) and $\varepsilon$ stands for the Levi-Civita symbol in two dimensions ($\varepsilon=i\tau_2$).
The letters $i,j=1,2,3$ denote flavour indices, while $a,b,c=1,2$ are SU(2)-gauge-group-related indices.
}
\end{table}
However, as it was motivated in our previous work~\cite{Husek:2020fru}, we consider an egalitarian flavour structure in the quark sector: equal entries for all quark flavours in the Wilson coefficient matrices $C_{ij}$ of the SMEFT (see below), where $i, j$ run over all quark flavours.
Hence, our LQ Yukawas are quark-flavour-blind, whereas in the lepton sector we allow for flavour violation --- although only in the third family, while keeping flavour universality for the first and second families.
The LQ Yukawa couplings in Table~\ref{tab:LQ}, namely $Y$ and $X$, will be assumed to be real.
Our ultra-violet (UV) Lagrangian, at the leptoquark mass scale, is then given by:
\begin{equation} 
\label{eq:UV}
{\cal L}_\text{UV} \, = \, {\cal L}_\text{SM} \, + \, {\cal L}_\text{S} \, + \, {\cal L}_\text{V} \, + \, {\cal L}_\text{LQ--SM} \, , 
\end{equation}
with ${\cal L}_\text{LQ--SM}$ consisting of the operators from Table~\ref{tab:LQ}.
Among various possible additional interactions, the so-called diquark couplings (i.e.\ when the LQ is coupled to a quark--antiquark pair) may appear at tree level in the Lagrangian \eqref{eq:UV} (although there are no analogous dilepton couplings to leptoquarks).
This entails a possible danger to matter stability, or, in turn, strong bounds on LQ masses or couplings.
Therefore, to avoid dealing with the proton-decay issue and since, in any case, diquark couplings do not play any role at tree level in CLFV processes, without any loss of generality we do not consider these couplings in this work and we have not included them in Table~\ref{tab:LQ}.
Although this precludes proton decay in the minimal scenario when assuming (besides the SM content) only one leptoquark species at a time, note that regardless of the presence or absence of diquark couplings, a richer scenario as the one treated here with all leptoquarks considered simultaneously is much more involved \cite{Dorsner:2016wpm,Crivellin:2021ejk}.
Hence, since these kinds of interactions do not affect our CLFV analysis and in order not to dive into this issue any further here, we simply assume, in what follows, that the proton is stable.

\section{The integration of leptoquarks}
\label{sec:EFT}

The aim of this work is to translate the bounds on the ratio $C/\Lambda^{2}_{\text{CLFV}}$ (containing the Wilson coefficients ($C$s) of the $D=6$ operators in the SMEFT and the high-energy scale $\Lambda_{\text{CLFV}}$) obtained by analysing charged-lepton-flavour-violating $\tau$ processes in Ref.~\cite{Husek:2020fru}, into constraints on the couplings and mass scales of the leptoquark Lagrangian described in Section~\ref{sec:LQLagrangian}.
\par 
Direct searches of leptoquarks have been extensively carried out at the LHC \cite{ParticleDataGroup:2020ssz}.
Lower bounds on their masses depend crucially on their spin (scalar or vector), their weak charges and generations involved, and the supposedly dominant decay products.
Generically, we can say that the present status requires $M_\text{LQ} >$ 1--2\,TeV, with a slight preference for the higher value \cite{ATLAS:2021oiz,CMS:2018qqq}.
Meanwhile, indirect determinations from the $B$ anomalies or lepton-number-violating processes \cite{ParticleDataGroup:2020ssz} require heavier LQs with masses of at least few TeVs (for ${\cal O}(1)$ LQ couplings).
Hence, it is rather fair to say that $M_\text{LQ} \gg \Lambda_\text{EW}$ and that their contribution to $D=6$ SMEFT monomials is hidden into the Wilson coefficients we denote $C$ (or, more generally, into $C/\Lambda^2$).
Assuming, in consequence, that there is such a mass gap between the SM particles and leptoquarks, we can integrate out the (heavy) leptoquark fields to recover the associated $D=6$ operators that contribute to the CLFV processes we study.
\par 
In our previous work \cite{Husek:2020fru}, we concluded that the strongest bounds from CLFV processes involving the tau lepton, namely hadronic tau decays, were imposed on the four-fermion operators and the operator responsible for the radiative decay $\tau \rightarrow \ell \gamma$; see Table~\ref{tab:D6} for a detailed list of these operators.
\begin{table}[!t]
\capstart
\begin{center}
\renewcommand{\arraystretch}{1.5}
\begin{tabular}{c|c||c|c}
\toprule
WC & Operator & WC & Operator \\
\midrule
 $C_{LQ}^{(1)}$ & $\left( \bar{L}_p \gamma_{\mu} L_r \right) \left( \bar{Q}_s \gamma^{\mu} Q_t \right)$   &
 $C_{LQ}^{(3)}$ & $\left( \bar{L}_p \gamma_{\mu} \tau^I L_r \right) \left( \bar{Q}_s \gamma^{\mu} \tau^I Q_t \right)$   \\
 $C_{eu}$ & $\left( \bar{e}_p \gamma_{\mu} e_r \right) \left( \bar{u}_s \gamma^{\mu} u_t \right)$  &
 $C_{ed}$ &  $\left( \bar{e}_p \gamma_{\mu} e_r \right) \left( \bar{d}_s \gamma^{\mu} d_t \right)$ \\
 $C_{Lu}$ & $\left( \bar{L}_p \gamma_{\mu} L_r \right) \left( \bar{u}_s \gamma^{\mu} u_t \right)$ &
 $C_{Ld}$ &  $\left( \bar{L}_p \gamma_{\mu} L_r \right) \left( \bar{d}_s \gamma^{\mu} d_t \right)$ \\ 
 $C_{Qe}$ &  $\left( \bar{Q}_p \gamma_{\mu} Q_r \right) \left( \bar{e}_s \gamma^{\mu} e_t \right)$ &
 $C_{LedQ}$ & $\left( \bar{L}^j_p e_r \right) \left( \bar{d}_s Q^j_t \right)$ \\ 
 $C_{LeQu}^{(1)}$ & $\left( \bar{L}_p^j e_r \right) \varepsilon_{jk}  \left( \bar{Q}_s^k u_t \right)$ &
 $C_{LeQu}^{(3)}$ & $\left( \bar{L}_p^j \sigma_{\mu \nu} e_r \right)  \varepsilon_{jk}  \left( \bar{Q}_s^k \sigma^{\mu \nu} u_t \right)$  \\
\midrule
 $C_{e W}$ &  $\left(\bar{L}_p \sigma^{\mu \nu} e_r \right)   \, \tau_I \, \varphi \, W_{\mu \nu}^I$ &
 $C_{eB}$ & $\left(\bar{L}_p \sigma^{\mu \nu} e_r \right)   \,  \varphi \, B_{\mu \nu}$ \\
\bottomrule
\multicolumn{2}{c}{}
\end{tabular}
\end{center}
\vspace*{-1cm}
\caption{\label{tab:D6}
Dominant $D=6$ SMEFT operators contributing to the CLFV processes generated by leptoquarks.
The notation (up to small apparent changes) is the one from Ref.~\cite{Grzadkowski:2010es}.
For the family indices, we use $p$, $r$, $s$ and $t$, while $j$ and $k$ are weak isospin indices.
For $I=1,2,3$, $\tau_I$ are the Pauli matrices, with $\varepsilon = i \tau_2$, and $\sigma^{\mu\nu}\equiv\frac i2[\gamma^\mu,\gamma^\nu]$.
In the first row, $\Lambda$ denotes the scale where the new dynamics arises.
The operators share the same notation with the associated couplings, substituting simply $C \rightarrow {\cal O}$, i.e.\ ${\cal O}_{LQ}^{(1)}$ and so on.
Four-fermion operators are obtained by integrating out, at the leading tree-level contribution, the leptoquark Lagrangian that we described in Section~\ref{sec:LQLagrangian}.
The last two operators in the table generate ${\cal O}_{\gamma} = c_\text{W} {\cal O}_{eB} - s_\text{W} {\cal O}_{eW}$, which is obtained by integrating out the leptoquarks at the leading one-loop contribution.}
\end{table}
Indeed, it was the ${\cal O}_{\gamma} = c_\text{W} {\cal O}_{eB} - s_\text{W} {\cal O}_{eW}$ operator (here and below, $c_\text{W} = \cos \theta_\text{W}$ and $s_\text{W} = \sin \theta_\text{W}$, with $\theta_\text{W}$ being the weak mixing (Weinberg) angle; see also Section~\ref{sec:DictionaryCLFV}) which was getting the strongest bound and has two relevant features.
Firstly, it incorporates a Higgs field (as can be seen in Table~\ref{tab:D6}).
Secondly, it is generated at the one-loop level upon leptoquark integration once the spontaneous symmetry breaking of the electroweak symmetry has taken place.
\par
In this article, we will consider the leptoquark contributions to both the four-fermion and ${\cal O}_\gamma$ operators at leading order.
We thus perform the low-energy matching ($M_\text{LQ} \gg \Lambda_\text{EW}$) of the leptoquark Lagrangian ${\cal L}_\text{UV}$ from Eq.~\eqref{eq:UV} with the SM extended by the $D=6$ operators (i.e.\ with the SMEFT Lagrangian) and obtain the relation among the Wilson coefficients and the couplings of the leptoquark framework --- after their running with the scale is taken into account.

\subsection{Four-fermion operators}
\label{ssec:4fo}

We start first with the matching giving the four-fermion CLFV operators.
In the classical limit (tree level) this can be achieved by using the equations of motion of the integrated fields, as follows from the application of the steepest descent method to determine the path integral of the effective action \cite{Boulware:1968zz,Koshkarov:1995al}.
This procedure gives a non-local Lagrangian.
Assuming that these scalars (mass $M_S$) and vectors (mass $M_V$) are very heavy, in comparison with the energy scale of the effective action, we can make an expansion in momenta in the corresponding solutions
for scalars $S$ and vectors $V$, producing a local action.
Sticking to the first order, we have, in a generic notation,
\begin{alignat}{4}
S_d
&\simeq  \frac{Y_{d,rs}^{\chi_1 \chi_2}}{M_S^2}\,\bar{\psi}_{\chi_1}^{\prime\,s} \, \psi_{\chi_2}^r\,,\qquad \qquad \qquad 
&
S_d^{\dagger}
&\simeq\frac{Y_{d,rs}^{\chi_1 \chi_2}}{M_S^2}\,\bar{\psi}_{\chi_1}^r\psi_{\chi_2}^{\prime\,s}
\,, \\
V_d^{\mu}
&\simeq-\frac{X_{d,rs}^{\chi_1 \chi_2}}{M_V^2}\,\bar{\psi}^{\prime\,s}_{\chi_1}\gamma^{\mu}\psi^r_{\chi_2} \,, \qquad 
&
V_d^{\mu \dagger}
&\simeq-\frac{X_{d,rs}^{\chi_1 \chi_2}}{M_V^2}\,\bar{\psi}^r_{\chi_1}\gamma^{\mu}\psi^{\prime\,s}_{\chi_2} 
\,,
\end{alignat}
where $d$ refers to the SU(2) representation (dimensionality) of the LQ field, and repeated indices (chiralities $\chi_k$ and flavours $r,s$) are summed over.
We will assume, in the following, that all the scalar leptoquarks, independently of their SU(2) quantum numbers, have the same mass $M_\text{S}$, and analogously for vector leptoquarks (having mass $M_\text{V}$).
Inserting these relations back into Eq.~\eqref{eq:UV} and introducing the notation of Table~\ref{tab:LQ} used in the rest of the paper, we obtain contributions to the effective Lagrangians accounting for effects stemming from the interactions of the scalar and vector LQ fields:
\begin{equation}
\begin{split}
\mathcal{L}_\text{S}^\text{eff}
& \, \supset \, 
\frac{Y_{d,ij}^{\chi_1\chi_2}Y_{d,mn}^{\chi_3\chi_4}}{M_\text{S}^2}\,
(\bar\psi_{\chi_1}^i\psi_{\chi_2}^{\prime\,j})
(\bar\psi_{\chi_4}^{\prime\,n}\psi_{\chi_3}^m)\,,
\qquad \\
\mathcal{L}_\text{V}^\text{eff}
&\, \supset \, 
\frac{X_{d,ij}^{\chi_1\chi_2}X_{d,mn}^{\chi_3\chi_4}}{M_\text{V}^2}\,
(\bar\psi_{\chi_1}^i\gamma_{\mu}\psi_{\chi_2}^{\prime\,j})
(\bar\psi_{\chi_4}^{\prime\,n}\gamma^{\mu}\psi_{\chi_3}^m)\,.
\end{split}
\end{equation}
\par 
Using the above prescription and restoring the SU(2) gauge structures, we end up with a list of effective $D = 6$ four-fermion operators.
In order to recognise the couplings of the SMEFT Lagrangian in Table~\ref{tab:D6} (or the modified basis suitable for the numerical analysis from Ref.~\cite{Husek:2020fru}), we have employed Fierz reordering and several relations and identities, as detailed in the Appendix~\ref{app:identities}.
Hence, considering generically $\Lambda=M_\text{LQ}$ for every leptoquark type,%
\footnote{As it will be clear further below, we consider two separate cases taking $\Lambda=M_\text{S}$ ($\Lambda=M_\text{V}$) for scalar (vector) leptoquarks.}
we can identify products of two LQ couplings with the Wilson coefficients.
The results are collected in Table~\ref{tab:C}.
\begin{table}[t]
\small
\centering
\renewcommand{\arraystretch}{1.5}
\setlength{\tabcolsep}{2pt}
\scalebox{0.68}{
\begin{tabular}[h]{c | c c c c c c c c c c}
\toprule
{ }LQ{ } & $C_{LQ}^{(1),\A\B\C\D}$ & $C_{LQ}^{(3),\A\B\C\D}$ & $C_{LeQu}^{(1),\A\B\C\D}$ & $C_{LeQu}^{(3),\A\B\C\D}$ & $C_{Qe}^{\A\B\C\D}$ & $C_{Lu}^{\A\B\C\D}$ & $C_{Ld}^{\A\B\C\D}$ & $C_{eu}^{\A\B\C\D}$ & $C_{ed}^{\A\B\C\D}$ & $C_{LedQ}^{\A\B\C\D}$ \\
\midrule
$S_3$ & $+\frac34 Y_{3,{\D}{\B}}^\text{LL}Y_{3,{\C}{\A}}^\text{LL}$ & $+\frac14 Y_{3,{\D}{\B}}^\text{LL}Y_{3,{\C}{\A}}^\text{LL}$ & \ts & \ts & \ts & \ts & \ts & \ts & \ts & \ts \\
$R_2$ & \ts & \ts & $-\frac12 Y_{2,{\D}{\A}}^\text{RL}Y_{2,{\B}{\C}}^\text{LR}$ & $-\frac18 Y_{2,{\D}{\A}}^\text{RL}Y_{2,{\B}{\C}}^\text{LR}$ & $-\frac12 Y_{2,{\C}{\B}}^\text{LR}Y_{2,{\D}{\A}}^\text{LR}$ & $-\frac12 Y_{2,{\C}{\B}}^\text{RL}Y_{2,{\D}{\A}}^\text{RL}$ & \ts & \ts & \ts & \ts \\
$\tilde{R}_2$ & \ts & \ts & \ts & \ts & \ts & \ts & $-\frac12 \tilde{Y}_{2,{\C}{\B}}^\text{RL}\tilde{Y}_{2,{\D}{\A}}^\text{RL}$ & \ts & \ts & \ts \\
$\tilde{S}_1$ & \ts & \ts & \ts & \ts & \ts & \ts & \ts & \ts & $+\frac12 \tilde{Y}_{1,{\D}{\B}}^\text{RR}\tilde{Y}_{1,{\C}{\A}}^\text{RR}$ & \ts \\
$S_1$ & $+\frac14 Y_{1,{\D}{\B}}^\text{LL}Y_{1,{\C}{\A}}^\text{LL}$ & $-\frac14 Y_{1,{\D}{\B}}^\text{LL}Y_{1,{\C}{\A}}^\text{LL}$ & $+\frac12 Y_{1,{\C}{\A}}^\text{LL}Y_{1,{\D}{\B}}^\text{RR}$ & $-\frac18 Y_{1,{\C}{\A}}^\text{LL}Y_{1,{\D}{\B}}^\text{RR}$ & \ts & \ts & \ts & $+\frac12 Y_{1,{\D}{\B}}^\text{RR}Y_{1,{\C}{\A}}^\text{RR}$ & \ts & \ts \\
\midrule
$U_3$ & $-\frac32 X_{3,{\C}{\B}}^\text{LL}X_{3,{\D}{\A}}^\text{LL}$ & $+\frac12 X_{3,{\C}{\B}}^\text{LL}X_{3,{\D}{\A}}^\text{LL}$ & \ts & \ts & \ts & \ts & \ts & \ts & \ts & \ts \\
$V_2$ & \ts & \ts & \ts & \ts & $+X_{2,{\B}{\D}}^\text{LR}X_{2,{\A}{\C}}^\text{LR}$ & \ts & $+X_{2,{\D}{\B}}^\text{RL}X_{2,{\C}{\A}}^\text{RL}$ & \ts & \ts & $-2 X_{2,{\C}{\A}}^\text{RL}X_{2,{\D}{\B}}^\text{LR}$ \\
$\tilde{V}_2$ & \ts & \ts & \ts & \ts & \ts & $+\tilde{X}_{2,{\D}{\B}}^\text{RL}\tilde{X}_{2,{\C}{\A}}^\text{RL}$ & \ts & \ts & \ts & \ts \\
$\tilde{U}_1$ & \ts & \ts & \ts & \ts & \ts & \ts & \ts & $-\tilde{X}_{1,{\C}{\B}}^\text{RR}\tilde{X}_{1,{\D}{\A}}^\text{RR}$ & \ts & \ts \\
$U_1$ & $-\frac12 X_{1,{\C}{\B}}^\text{LL}X_{1,{\D}{\A}}^\text{LL}$ & $-\frac12 X_{1,{\C}{\B}}^\text{LL}X_{1,{\D}{\A}}^\text{LL}$ & \ts & \ts & \ts & \ts & \ts & \ts & $-X_{1,{\C}{\B}}^\text{RR}X_{1,{\D}{\A}}^\text{RR}$ & $+2 X_{1,{\D}{\A}}^\text{LL}X_{1,{\C}{\B}}^\text{RR}$ \\
\bottomrule
\end{tabular}
}
\caption{Results of the matching of pairs of Yukawa couplings stemming from each leptoquark type listed in Table~\ref{tab:LQ} to the Wilson coefficients of the four-fermion operators of SMEFT from Table~\ref{tab:D6}.
Notice that, owing to the Fierz rearrangement, we also obtain contributions containing tensorial operators, i.e.\ $C_{LeQu}^{(3)}$.
\label{tab:C}
}
\end{table}
\par 
Leptoquarks can mediate lepton-flavour violation even when all Yukawa couplings are considered equal.
However, since we have motivated and assumed an enhancement of this phenomenon for the third lepton family (see Section~\ref{sec:intro}), we consider the $\tau$-related Yukawas different (potentially larger) than those for the other two charged leptons, so that the (potentially stronger) limits imposed (from other works) on the first- or second-family-related Yukawas do not apply to our case.
This fact, together with the flavour considerations (quark-flavour-blind Yukawas) discussed in the previous section, entail a rather simple flavour structure for the Yukawa matrices.
In general,
\begin{equation}
\label{eq::YukawaStructure}
Y_{d,rs}^{\chi_1\chi_2}=
\begin{pmatrix}
\, y_{d}^{\chi_1\chi_2} & y_{d}^{\chi_1\chi_2} & y_{d\,\tau}^{\chi_1\chi_2} \, \vspace{2mm} \\
\, y_{d}^{\chi_1\chi_2} & y_{d}^{\chi_1\chi_2} & y_{d\,\tau}^{\chi_1\chi_2} \, \vspace{2mm} \\
\, y_{d}^{\chi_1\chi_2} & y_{d}^{\chi_1\chi_2} & y_{d\,\tau}^{\chi_1\chi_2} \, \\
\end{pmatrix}_{rs},\qquad \qquad
\,y_{d}^{\chi_1\chi_2} \neq y_{d\,\tau}^{\chi_1\chi_2}\,.
\end{equation}
The only exception to this prescription is, by definition, for $Y_2^\text{LR}$, the structure of which would be the $Y_{d}^{\chi_1\chi_2}$ from above transposed.
For vector leptoquarks, we proceed analogously taking $Y \rightarrow X$.

\subsection{Dipole operator and \texorpdfstring{\boldmath $C_{\gamma}$}{Cgamma}}
\label{ssec:tff}

Among all the $D=6$ operators present in the relevant CLFV basis of Ref.~\cite{Grzadkowski:2010es}, we concluded that the dominant contribution to the processes that we analysed \cite{Husek:2020fru} stemmed from dipole operators via the exchange of a photon, namely ${\cal O}_{\gamma} = c_\text{W} {\cal O}_{eB} - s_\text{W} {\cal O}_{eW}$ as written in terms of the operators from Table~\ref{tab:D6}.
However, within the leptoquark framework, the leading-order contribution to this operator occurs at one loop.
In spite of that, and given the relevance of the strong bound on the corresponding WC, we think that it is pertinent that we also consider the leptoquark contribution to this coupling.
\par 
We will consider the process $\ell_1 \rightarrow \ell_2 \gamma$, with $\ell_1=\tau$ and $\ell_2=e,\mu$; the detailed analysis can be found in Appendix~\ref{app:CgammaMatching}.
Note that for the CLFV processes under consideration, a leptoquark and a quark enter necessarily in the loop.
The relevant leptoquarks are thus $S_{3}^{1/3}$, $R_{2}^{5/3}$ and $S_{1}^{1/3}$ (with the superscripts identifying the respective electric charges) but only the last two provide a dominant contribution.
Indeed, the main contribution to these amplitudes comes from chirality-enhanced effects.
The triplet scalar LQ only couples to fermion doublets --- hence to two fields of the same (left) chirality (see Table~ \ref{tab:LQ}) --- and their amplitudes are thus suppressed by the mass of either {\em lepton} in the process \cite{Lavoura:2003xp}.
Meanwhile, the singlet and doublet LQs couple to both chiralities, and their amplitudes are hence proportional to the mass of the {\em quark} in the loop.
Accordingly, the dominant amplitudes will be those which include the exchange of a LQ and a top quark.
As a consequence, we will only consider the third-family quark.
In addition --- and this is also further motivated from the Yukawa flavour structure of Eq.~\eqref{eq::YukawaStructure} --- all quarks couple with the same strength to leptoquarks and leptons, the difference among the latter then stemming from the specific lepton flavour only.
Related to the dipole operator, we will thus only present the bounds for these singlet and doublet LQs, assuming that --- as it also happens for the other LQs --- the Yukawas of $S_{3}^{1/3}$ are more strongly constrained from the four-fermion operators (as it turns out to be the case).
Note that the rest of the leptoquarks would also contribute through their couplings to other quark flavours, e.g.\ to bottom and charm quarks.
Accordingly, their Yukawas would also be constrained by the $C_\gamma$ bound.
However, similar arguments as for $S_{3}^{1/3}$ apply.
\par 
Finally, since we are considering the $\ell_{1} \to \ell_{2} \gamma$ process to perform the aforementioned matching, we can take into account also the bounds provided by direct searches for $\tau \to \ell \gamma$, with $\ell=e,\mu$~\cite{HFLAV:2019otj, Kou:2018nap}.
For this we use the results from Ref.~\cite{Lavoura:2003xp} and compare both types of constraints in Section~\ref{sec:ScalarLeptoquarks};
see also Ref.~\cite{Dedes:2021abc} for a complete matching of the scalar leptoquarks at the one-loop level to operators up to $D=6$.

\section{Results}
\label{sec:Results}

We address now the results of our analyses.
The constraints for the SMEFT $D=6$ Wilson coefficients found in the previous work~\cite{Husek:2020fru} were obtained considering the expected sensitivity of the NA64 experiment at CERN~\cite{Gninenko:2018num} (for $\ell$--$\tau$ conversion in nuclei), and the current and expected results of Belle~\cite{HFLAV:2019otj} and Belle II~\cite{Kou:2018nap} experiments (for hadronic $\tau$ decays).
We arrived at the final numerical values for the bounds by means of the open-source tool \texttt{HEPfit} \cite{DeBlas:2019ehy}.
These have to be now translated to the leptoquark-associated ratios $yy^{\prime}/M_\text{S}^2$ and $xx^{\prime}/M_\text{V}^2$, which is done as explained in appendix~\ref{app:limits}.
Let us list some important aspects of relating the SMEFT and leptoquark frameworks.
\par 
In Ref.~\cite{Husek:2020fru}, all $D=6$ operators contributing to CLFV processes were considered simultaneously, and it was shown that a naive single-operator analysis would result in overestimated (stronger) constraints on the corresponding WCs, simply due to the lack of correlations among the operators.
On a similar basis, we think that it is more natural to consider a scenario in which not only one type of scalar or vector LQ drives the CLFV dynamics.
Furthermore, the limits extracted by reviewing the results on the processes involving the CLFV phenomena in our previous work assumed only one energy scale $\Lambda_\text{CLFV}$.
On the other hand, within the leptoquark framework we have many possible scales --- LQ masses.
The nature of different LQs (scalar and vector) can, in principle, be completely unrelated, and stem from a very different origin and/or scale.
Hence, we will divide our analysis into two separate parts:
We consider first the simultaneous contribution of all different types of scalar leptoquarks (described by a common mass $M_\text{S}$) and then of all vector leptoquarks (common mass $M_\text{V}$), with $M_\text{S} \neq M_\text{V}$ in general.
\par 
In our previous work~\cite{Husek:2020fru}, we extracted bounds on the aforementioned $D=6$ SMEFT Wilson coefficients from CLFV $\tau$-involved processes within two distinct scenarios: First, we considered new physics driving only CLFV in the $\tau$ sector. Second, on top of these phenomena, we also allowed for flavour-changing neutral currents (FCNC) in the quark sector --- transitions $(\bar{c}u)$, $(\bar{b}s)$, $(\bar{s}d)$, etc. However, for the second scenario, we made the assumption that the energy scales $\Lambda_{\text{CLFV}}$ and $\Lambda_{\text{FCNC}}$ driving these phenomena were the same. Within the leptoquark framework, due to the nature of the LQ--lepton--quark interactions, this hypothesis is much better motivated, i.e.\ leptoquarks readily mediate FCNC processes even in the presence of negligible off-diagonal Yukawa couplings.
Hence, we present all the bounds obtained in the (second) FCNC case.
\par
Finally, since the constraints are obtained from CLFV $\tau$-involved processes, we performed the running of all the Wilson coefficients to the scale given by the $\tau$ mass; see Ref.~\cite{Husek:2020fru} for more details.
Accordingly, the bounds on the pairs of Yukawa couplings are presented at this scale.

\subsection{Dictionary for the CLFV effective basis}
\label{sec:DictionaryCLFV}

In Table~\ref{tab:C}, we collect the results for the matching of the leptoquark framework to the SMEFT operators listed in Table~\ref{tab:D6} upon tree-level integration
of the leptoquarks.
However, the basis of $D=6$ operators presented in Table~\ref{tab:D6} does not exactly match the basis constrained from the CLFV processes studied in Ref.~\cite{Husek:2020fru}.
A short dictionary is given here to ease the transition.
\par 
Consistently with Ref.~\cite{Husek:2020fru}, we only consider the QCD running of the SMEFT operators.
On top of that, we do not consider the (very suppressed) QCD running of the vector and axial-vector currents, while
the running of the scalar densities is related to the divergences of the vector currents through, for instance,
$\partial^{\mu} (\overline{s} \gamma_{\mu} u) = i (m_s-m_u) (\overline{s}u)$. 
Hence, the scalar Wilson coefficients were redefined into scale-invariant $C_{LedQ}^{\prime}$ and $C_{LeQu}^{(1)\, \prime}$, such that
\begin{equation}
C_{LeQu}^{(1)}=\frac{m_{i}}{m_{\tau}}\,C_{LeQu}^{(1)\,\prime}
\,,\qquad\qquad
C_{LedQ}=\frac{m_{i}}{m_{\tau}}\,C_{LedQ}^{\prime}
\,,
\end{equation}
with $m_{i}$ being the mass of the quark involved in the process.
For the related bound, this just entails a change by a factor proportional to $m_{i}$.
For $u$-processes ($C_{LeQu}^{(1)}$) in $\ell$--$\tau$ conversion, since we considered $m_{u}=0$, the overall factor becomes $m_{c}$; in hadronic $\tau$ decays though, despite having taken into account the $\chi$PT scale-independent combination $2B_{0}M_{q}\simeq \mathcal{M}_\text{P}$ (with $M_{q}$ being the diagonal matrix of the light-quark masses and $\mathcal{M}_\text{P}$ the physical-mass matrix of the pseudoscalar Goldstone bosons), we will consider the mass of the up quark at the energy scale of the $\tau$ mass.
For $d$-processes ($C_{LedQ}$), the situation is slightly more involved:
For the lightest quark in $\ell$--$\tau$ conversion, it holds accordingly $m_{d}=0$.
However, for $s$ and $b$ quarks, even though different masses were considered, only a single bound was obtained, introducing an ambiguity when translating the constraints on the primed WCs into the non-primed.
This is solved for vector leptoquarks%
\footnote{Note that the scalars do not contribute at tree level to $C_{LedQ}$, as can be seen in Table~\ref{tab:C}.} %
by considering the most conservative bound, i.e.\ taking $m_{i}=m_{b}$.
For hadronic $\tau$ decays, related accordingly to the most conservative bound, $m_s$ at the scale of $m_{\tau}$ is used.
\par 
Considering now the $\gamma$ dipole case, the $C_{eB}$ and $C_{e W}$ (see Table~\ref{tab:D6}) were rotated into $C_{\gamma}$ and $C_{Z}$, parameterising the $\gamma$- and $Z$-mediated contributions separately, i.e.
\begin{equation}
\label{eq:RedefinitionDipoleWC}
\begin{pmatrix}
C_{\gamma} \\
C_{Z}
\end{pmatrix}=
\begin{pmatrix}
c_\text{W} & -s_\text{W}\\
s_\text{W} & c_\text{W}
\end{pmatrix}
\begin{pmatrix}
C_{eB} \\
C_{eW}
\end{pmatrix}.
\end{equation}
However, neither the $C_Z$ nor the rest of $Z$-mediated contributions (provided by the coupling of the $Z$ to leptoquarks) will be considered in our analyses, since the related leading LQ contribution appears at one loop and the resulting constraints would thus not be competitive.
\par 
The absence of some operators within each leptoquark scenario
poses another caveat related to the correlations between the present and absent (or numerically negligible) operators.
The correlations found in Ref.~\cite{Husek:2020fru} tend to relax the bounds set on individual Wilson coefficients:
A loss of any correlation would turn into stronger constraints for the specific WC.
In this study, when the results are presented below, in case there is some significant correlation between present and absent WCs, we will adopt the most conservative view and assume that, somehow, the correlated operators could be present and the weakest constraint is used.
To explain this point better, let us provide an example.
We found that, for instance, (absent) $C_{\varphi L}^{(1)\, \prime}$ correlates strongly with $C_{LQ}^{(1)}$, $C_{LQ}^{(3)}$ and $C_{Lu}$ (see Fig.~10 in Ref.~\cite{Husek:2020fru}), and that the limits on the three latter mentioned stem from the global analysis where these correlations are taken into account.
The same happens for the correlation of the rotated $C_{Z}$ with $C_{LeQu}^{(3)}$ and $C_{\gamma}$, the latter of which further translates into $C_{eB}$ and $C_{eW}$.
To be more specific, let us discuss other examples within the vector LQ scenario:
$C_{LedQ}$ is strongly correlated with $C_{e\varphi}$, although the latter is omitted here; $C_{LeQu}^{(1)}$ and $C_{LeQu}^{(3)}$ are completely absent and, while the first is practically uncorrelated, the second correlates significantly with the (rotated) $C_{\gamma}$ (considered here) and $C_{Z}$ (not taken into account).
\par 
The previous paragraph refers to the case of $\tau$ decays; for $\ell$--$\tau$ conversion in nuclei, the only Wilson coefficients used here are either correlated among each other ($C_{LeQu}^{(1)\, \elltau}$ and $C_{LeQu}^{(3)\, \elltau}$) or not correlated with other WCs ($C_{LedQ}^{\elltau}$).
\par
In summary, the correlations between the Wilson coefficients of the SMEFT are transferred into the corresponding constraints on the products of LQ Yukawa-like couplings.
As commented above, we have taken into account this feature in our analysis and we consider the bounds we present conservative.

\subsection{Scalar leptoquarks}
\label{sec:ScalarLeptoquarks}

We have considered the contribution of all scalar leptoquarks to both the four-fermion and the dipole operators.

\subsubsection{Four-fermion constraints}

For scalar leptoquarks, assuming a common energy scale $\Lambda_\text{CLFV}=M_\text{S}$, one can unambiguously relate pairs of Yukawa LQ couplings with distinct linear combinations of Wilson coefficients.
We find
\begin{alignat}{2}
y_3^\text{LL}y_{3\,\tau}^\text{LL}&=C_{LQ}^{(1)}+C_{LQ}^{(3)}\,,\notag\\
y_2^\text{RL}y_{2\,\tau}^\text{RL}&=-2C_{Lu}\,,
&y_2^\text{LR}y_{2\,\tau}^\text{LR}&=-2C_{Qe}\,,\notag\\
y_1^\text{LL}y_{1\,\tau}^\text{LL}&=C_{LQ}^{(1)}-3 C_{LQ}^{(3)}\,,
&y_1^\text{RR}y_{1\,\tau}^\text{RR}&=2C_{eu}\,,\notag\\
\tilde{y}_2^\text{RL}\tilde{y}_{2\,\tau}^\text{RL}&=-2C_{Ld}\,,
&\tilde{y}_1^\text{RR}\tilde{y}_{1\,\tau}^\text{RR}&=2C_{ed}\,,\label{eq:YY}\\
y_{2\,\tau}^\text{RL}y_2^\text{LR}&=-C_{LeQu}^{(1)\,\elltau}-4 C_{LeQu}^{(3)\,\elltau}\,,\qquad\qquad
&y_2^\text{RL}y_{2\,\tau}^\text{LR}&=-C_{LeQu}^{(1)\, \tauh}-4 C_{LeQu}^{(3)\, \tauh}\,,\notag\\
y_{1\,\tau}^\text{LL}y_1^\text{RR}&=C_{LeQu}^{(1)\,\elltau}-4 C_{LeQu}^{(3)\, \elltau}\,,
&y_1^\text{LL}y_{1\,\tau}^\text{RR}&=C_{LeQu}^{(1)\, \tauh}-4 C_{LeQu}^{(3)\,\tauh}\,.\notag
\end{alignat}
Here, the superscripts `${\elltau}$' and `${\tauh}$' stand for the respective processes ($\ell$--$\tau$ conversion in nuclei and hadronic $\tau$ decays), from which the bounds on WCs stem, given that the corresponding pairs of Yukawa couplings (which we simply refer to as `Yukawa pairs' further on) contribute only to one of those processes.
\par 
The numerical results are presented in Table~\ref{tab:Belle_BelleIIScalar}, which shows constraints for two different aspects of the leptoquark framework:
\begin{table}[tb]
\capstart
\begin{center}
\renewcommand{\arraystretch}{1.3}
\begin{tabularx}{0.98\textwidth}
{%
    >{\centering\arraybackslash}c %
  | >{\centering\arraybackslash}X %
    >{\centering\arraybackslash}X %
  | >{\centering\arraybackslash}X %
    >{\centering\arraybackslash}c  %
}
\toprule
$\tau$ decays &
\multicolumn{2}{c|}{Upper bounds on $\vphantom{\Bigg(}\displaystyle\frac{|yy^\prime|}{M_\text{S}^2}\,[10^{-3}\,\text{TeV}^{-2}]$} &
\multicolumn{2}{c}{Lower bounds on $M_\text{S}$\,[TeV]}\\
\hline
 Yukawa pair & Belle & Belle II & Belle & Belle II\\
\midrule
 $|y_3^\text{LL}y_{3\,\tau}^\text{LL}|$ & $12$ & $1.9$ & $9.1$ & $23$\\
 $|y_2^\text{RL}y_{2\,\tau}^\text{RL}|$ & $47$ & $5.0$ & $4.6$ & $14$\\
 $|y_2^\text{LR}y_{2\,\tau}^\text{LR}|$ & $17$ & $2.6$ & $7.8$ & $20$\\
$|y_2^\text{RL}y_{2\,\tau}^\text{LR}|$ & $28$ & $3.7$ & $6.0$ & $16$\\
$|\tilde{y}_2^\text{RL}\tilde{y}_{2\,\tau}^\text{RL}|,\,|\tilde{y}_1^\text{RR}\tilde{y}_{1\,\tau}^\text{RR}|$ & $20$ & $3.0$ & $7.1$ & $18$\\
$|y_1^\text{LL}y_{1\,\tau}^\text{LL}|$ & $64$ & $7.7$ & $3.9$ & $11$\\
$|y_1^\text{RR}y_{1\,\tau}^\text{RR}|$ & $34$ & $4.1$ & $5.4$ & $16$\\
$|y_1^\text{LL}y_{1\,\tau}^\text{RR}|$ & $28$ & $3.7$ & $6.0$ & $16$\\
\midrule
\midrule
$\ell$--$\tau$ conversion &
\multicolumn{2}{c|}{Upper bounds on $\vphantom{\Bigg(}\displaystyle\frac{|yy^\prime|}{M_\text{S}^2}\,[10^{0}\,\text{TeV}^{-2}]$} &
\multicolumn{2}{c}{Lower bounds on $M_\text{S}$\,[TeV]}\\
\hline
 Yukawa pair & $e$--$\tau$ & $\mu$--$\tau$ & $e$--$\tau$ & $\mu$--$\tau$\\
\midrule
$|y_{2\,\tau}^\text{RL}y_2^\text{LR}|$ & $350$ & $2.3$ & $0.054$ & $0.66$\\
$|y_{1\,\tau}^\text{LL}y_1^\text{RR}|$ & $250$ & $1.8$ & $0.063$ & $0.75$\\
\bottomrule
\end{tabularx}
\end{center}
\vspace*{-0.5cm}
\caption{\label{tab:Belle_BelleIIScalar}
Obtained bounds for the scalar leptoquark case from our results in Ref.~\cite{Husek:2020fru}.
In the left-hand part of the table, we present {\em upper} bounds on the ratio $|yy^\prime|/M_\text{S}^2$; these numbers, of course, also correspond to (with appropriate power of 10) {\em upper} bounds on the Yukawa pairs $|yy^{\prime}|$, assuming $M_\text{S}=1$\,TeV.
On the right, there are {\em lower} bounds on the probed energy scale of the scalar leptoquarks mediating CLFV phenomena ($\Lambda_\text{CLFV}=M_\text{S}$), considering $|yy^{\prime}|\approx1$.
The strongest bounds found are shown, most of which are stemming from the $\tau$ decays analysis (Belle and Belle II results), the exception being the last couple of rows dedicated to Yukawas contributing only to $\ell$--$\tau$ conversion.
The values are given at the 99.8\,\% confidence level.
}
\end{table}
{\em lower} bounds for the masses of scalar leptoquarks assuming the product of LQ Yukawa couplings being of ${\cal O}(1)$, and, in turn, {\em upper} bounds for the Yukawa pairs, taking $M_\text{S} = 1$\,TeV.
Note that the two Yukawa pairs $y_{2\,\tau}^\text{RL}y_2^\text{LR}$ and $y_{1\,\tau}^\text{LL}y_1^\text{RR}$ are unconstrained by $\tau$ decays.
Thus, the $\ell$--$\tau$ conversion limits are considered instead.
Regarding these last couple of bounds, the allowed values for the Yukawa pairs from $e$--$\tau$ conversion exceed by far the limits suggested by perturbativity considerations; on the other hand, the bounds on the probed mass (with $|yy^{\prime}|\approx1$) just indicate that the experimental constraints from the $\ell$--$\tau$ conversion in nuclei cannot yet compete with those stemming from the $\tau$ decays.

\subsubsection{Dipole operator and \texorpdfstring{\boldmath $C_{\gamma}$}{Cgamma} constraints}

In order to translate the constraint on the dipole WC $C_{\gamma}$ obtained in Ref.~\cite{Husek:2020fru} into the most general leptoquark framework considered in this work, we have evaluated the (leading) one-loop contribution to the $\ell_1 \rightarrow \ell_2 \gamma$ process within the leptoquark UV theory (\ref{eq:UV}) and, upon integrating out the leptoquark fields, performed the matching of the result with the one obtained within SMEFT; for details, see Appendix~\ref{app:CgammaMatching}.
As explained in Section~\ref{ssec:tff}, only two LQ fields contribute to this process, namely $S_1^{1/3}$ and $R_2^{5/3}$.
\par 
For completeness, we also consider the most stringent bounds stemming from the $\tau \to \ell \gamma$ direct searches (with  $\ell=e,\mu$) by Belle and BaBar \cite{HFLAV:2019otj}, and compare them with the constraints obtained from $C_{\gamma}$ based on Ref.~\cite{Husek:2020fru}, in the same way as it is explained for the single leptoquark scenarios below.
Since these processes provide bounds on the same ratios $yy^{\prime}/M_\text{S}^{2}$ as the $C_{\gamma}$ constraint, we compare all of them in Table~\ref{tab:CgammaLimitsR2_S1}.
\begin{table}[tb]
\capstart
\begin{center}
\renewcommand{\arraystretch}{1.5}
\begin{tabularx}{0.98\textwidth}
{%
    >{\centering\arraybackslash}c %
  | >{\centering\arraybackslash}X %
    >{\centering\arraybackslash}X %
  | >{\centering\arraybackslash}X %
   >{\centering\arraybackslash}c  %
}
\toprule
$C_{\gamma}/\Lambda_{\text{CLFV}}^{2}$ &
\multicolumn{2}{c|}{Upper bounds on $\vphantom{\Bigg(}\displaystyle\frac{|yy^\prime|}{M_\text{S}^2}\,[10^{-3}\,\text{TeV}^{-2}]$} &
\multicolumn{2}{c}{Lower bounds on $M_\text{S}$\,[TeV]}\\
\hline
Yukawa pair & Belle & Belle II & Belle & Belle II\\
\midrule
$|y_{2\,\tau}^\text{RL}y_2^\text{LR}|$ & $150$ & $19$ & $3.1$ & $8.6$\\
$|y_{1\,\tau}^\text{LL}y_1^\text{RR}|$ & $21$ & $2.7$ & $8.2$ & $23$\\
\midrule
\midrule
$\tau \to \ell \gamma$ &
\multicolumn{2}{c|}{Upper bounds on $\vphantom{\Bigg(}\displaystyle\frac{|yy^\prime|}{M_\text{S}^2}\,[10^{-3}\,\text{TeV}^{-2}]$} &
\multicolumn{2}{c}{Lower bounds on $M_\text{S}$\,[TeV]}\\
\hline
Yukawa pair & $\tau \to e \gamma$ & $\tau \to \mu \gamma$ & $\tau \to e \gamma$ & $\tau \to \mu \gamma$\\
\midrule
$|y_{2\,\tau}^\text{RL}y_2^\text{LR}|$ & $0.66$ & $0.79$ & $83$ & $75$\\
$|y_{1\,\tau}^\text{LL}y_1^\text{RR}|$ & $1.4$ & $1.7$ & $71$ & $64$\\
\bottomrule
\end{tabularx}
\end{center}
\vspace*{-0.5cm}
\caption{\label{tab:CgammaLimitsR2_S1}
Obtained bounds for the $R_{2}^{5/3}$ and $S_{1}^{1/3}$ single-scalar-leptoquark cases (only one scalar leptoquark considered at a time).
The values in the top part of the table stem from the bounds on $C_{\gamma}/ \Lambda_{\text{CLFV}}^{2}$ (once the four-fermion constraints are applied) based on our results from Ref.~\cite{Husek:2020fru}.
The bottom part shows bounds obtained from direct searches for $\tau \to \ell \gamma$ by Belle and BaBar experiments~\cite{HFLAV:2019otj}.
In the left-hand part of the table, we present {\em upper} bounds on the ratio $|yy^\prime|/M_\text{S}^2$; these numbers also correspond to {\em upper} bounds on the Yukawa pairs $|yy^{\prime}|$, assuming $M_\text{S}=1$\,TeV.
On the right, there are {\em lower} bounds on the probed energy scale of the scalar leptoquarks mediating CLFV phenomena ($\Lambda_{\text{CLFV}}=M_\text{S}$), considering $|yy^{\prime}|\approx1$.
The strongest bounds found on $C_{\gamma}$ are shown and are stemming from the $\tau$ decays analysis (i.e.\ from the Belle and Belle II results).
The values from $C_{\gamma}$ are given at the 99.8\,\% confidence level while bounds from direct searches are given at the 90\,\% confidence level.
}
\end{table}
\par
In the following, we explain the different cases that we analyse:
\myparagraph{Leptoquark \boldmath ${R_{2}^{5/3}}$}
In a framework with only the $R_{2}^{5/3}$ leptoquark, the main contribution to the process $\ell_{1}\to \ell_{2} \gamma$ --- once the LQ is integrated out --- provides the following matching between the SMEFT $\gamma$ dipole operator and the UV theory (see Appendix~\ref{app:CgammaMatching}):
\begin{equation}
\label{eq::MatchingR2Cgamma}
\bigg(\frac{C_{\gamma}}{\Lambda_{\text{CLFV}}^{2}}\bigg)^{2}
=\frac{e^{2} N_C^2 m_{t}^{2}V_{tb}^{2}}{2^{11}\pi^{4}v^{2}\MR^{4}}\Big( Q_{\hspace{-.25ex}R_{2}^{5/3}}-3 Q_{t} \Big)^{2}
\Big[ (y_{2 \, \tau}^\text{RL}y_{2}^\text{LR})^{2} + (y_{2 \, \tau}^\text{LR}y_{2}^\text{RL})^{2} \Big] \, .
\end{equation} 
Above, $m_t$ is the mass of the top quark, $v = \langle 0 | \varphi | 0 \rangle = (\sqrt{2} G_\text{F})^{-1/2}$, and $V_{tb}$ is the corresponding Cabibbo--Kobayashi--Maskawa matrix element.
\par 
The main bounds coming from the four-fermion operators on the above-appearing Yukawa pairs are $\big|y_{2 \, \tau}^\text{RL}y_{2}^\text{LR}\big| =8 \, C_{LeQu}^{(3)\, \elltau} \lesssim 1.1 \, \big(\frac{\Lambda_{\text{CLFV}}}{\text{TeV}}\big)^{2} $ and $\big|y_{2 \, \tau}^\text{LR}y_{2}^\text{RL}\big|=2 \, C_{LeQu}^{(1)\, \tauh} \lesssim 5.8 \times 10^{-3} \,\big(\frac{\Lambda_{\text{CLFV}}}{\text{TeV}}\big)^{2} $ (Belle).
Since the bound from $\tau$ decays constrains the value of $\big|y_{2 \, \tau}^\text{LR}y_{2}^\text{RL}\big|$ by about 3 orders of magnitude stronger than the bound from $\ell$--$\tau$ conversion does for $\big|y_{2 \, \tau}^\text{RL}y_{2}^\text{LR}\big|$, we can, in Eq.~\eqref{eq::MatchingR2Cgamma}, neglect the contribution from the former and use the limits from Belle and Belle II for $C_{\gamma}/ \Lambda_{\text{CLFV}}^{2}$ to constrain $\big|y_{2 \, \tau}^\text{RL}y_{2}^\text{LR}\big| / \MR^{2}$.
The results are presented in Table~\ref{tab:CgammaLimitsR2_S1}.
As we can see, the Yukawa pair $ \big|y_{2 \, \tau}^\text{RL}y_{2}^\text{LR}\big|$ as well as the corresponding probed scale $\MR$ both receive a stronger constraint than in the previous case (cf.\ Table~\ref{tab:Belle_BelleIIScalar}) where this pair was only sensitive to the limits from $\ell$--$\tau$ conversion in nuclei, and was, accordingly, bounded rather weakly.
\myparagraph{Leptoquark \boldmath ${S_{1}^{1/3}}$}
In this case, we end up with the following matching (see Appendix~\ref{app:CgammaMatching}):
\begin{equation}
\label{eq::MatchingS1Cgamma}
\bigg(\frac{C_{\gamma}}{\Lambda_{\text{CLFV}}^{2}}\bigg)^{2}
=\frac{e^{2} N_C^2 m_{t}^{2}V_{tb}^{2}}{2^{11}\pi^{4}v^{2}\MS^{4}}\Big(Q_{\hspace{-.25ex}S_{1}^{1/3}}-3 Q_{\bar{t}} \Big)^{2}
\Big[ (y_{1 \, \tau}^\text{LL}y_{1}^\text{RR})^{2} + (y_{1 \, \tau}^\text{RR}y_{1}^\text{LL})^{2} \Big] \, .
\end{equation} 
The main bounds coming from the four-fermion operators on the Yukawa pairs involved are $\big|y_{1 \, \tau}^\text{LL}y_{1}^\text{RR}\big|=8\,C_{LeQu}^{(3)\, \elltau} \lesssim 1.1\,\big(\frac{\Lambda_{\text{CLFV}}}{\text{TeV}}\big)^{2} $ and $\big|y_{1 \, \tau}^\text{RR}y_{1}^\text{LL}\big|=2 \, C_{LeQu}^{(1)\, \tauh} \lesssim 5.8 \times 10^{-3} \big(\frac{\Lambda_{\text{CLFV}}}{\text{TeV}}\big)^{2} $ (Belle).
As before, in Eq.~\eqref{eq::MatchingS1Cgamma}, we can thus neglect the contribution of $\big|y_{1 \, \tau}^\text{RR}y_{1}^\text{LL}\big|$ (which receives stronger bounds), and use the limits from Belle and Belle II on $C_{\gamma}/ \Lambda_{\text{CLFV}}^{2}$ to constrain $\big|y_{1 \, \tau}^\text{LL}y_{1}^\text{RR}\big| / \MS^{2}$.
The results are presented in Table~\ref{tab:CgammaLimitsR2_S1}.
Again the bound on the dipole operator helps to constrain the otherwise weakly-bounded (cf.\ Table~\ref{tab:Belle_BelleIIScalar}) Yukawa pair $\big|y_{1 \, \tau}^\text{LL}y_{1}^\text{RR}\big|$.
\par 
Finally, note that in both single-leptoquark cases --- even though the Yukawa pairs are more constrained from the bound on $C_\gamma$ stemming from $\tau$ decays than from $\ell$--$\tau$ conversion limits --- the probed energy scale (when assuming $|yy^{\prime}|\approx 1$) is still smaller than the value obtained from the four-fermion bound of 13\,TeV and 36\,TeV from Belle and Belle II limits, respectively, on $\big|y_{2 \, \tau}^\text{LR}y_{2}^\text{RL}\big|/\MR^{2}$ (for $R_{2}^{5/3}$) and $\big|y_{1 \, \tau}^\text{RR}y_{1}^\text{LL}\big|/\MS^{2}$ (for $S_{1}^{1/3}$).
This is due to the fact that the same mass enters for all WCs (operators).
\myparagraph{Leptoquarks \boldmath ${R_{2}^{5/3} + S_1^{1/3}}$}
When both contributing leptoquarks $R_{2}^{5/3}$ and $S_{1}^{1/3}$ are considered at the same time, the corresponding matching given in Eq.~\eqref{eq::FinalCgammaMatching} --- for natural values of the Yukawa pairs $|yy^\prime|\approx 1$ --- is probing $M_\text{S}\gtrsim 4.7$\,TeV stemming from the Belle $\tau$ decay limits (using $C_{\gamma}/\Lambda_{\text{CLFV}}^{2} \lesssim 7 \times 10^{-5} \, \mbox{TeV}^{-2}$) or $M_\text{S}\gtrsim 13$\,TeV stemming from the Belle II limits (using $C_{\gamma}/\Lambda_{\text{CLFV}}^{2} \lesssim 9 \times 10^{-6} \, \mbox{TeV}^{-2}$).
However, it does not provide relevant bounds on the Yukawa pairs.

\subsection{Vector leptoquarks}
\label{sec:VectorLeptoquarks}

We will consider the contribution of vector leptoquarks to the four-fermion operators only.
Upon their integration, with $\Lambda_\text{CLFV}=M_\text{V}$, we end up with 11 distinct Yukawa pairs contributing to 8 Wilson coefficients (see Table~\ref{tab:C}).
Due to the flavour structure of the Yukawas and the integration of the LQs themselves, most of the Yukawa pairs contribute equally to both the $\ell$--$\tau$ conversion and $\tau$ decays.
However, there are two pairs (stemming from the last column in Table~\ref{tab:C}) which are not symmetric (built from two same matrices) and produce couplings in combinations relevant only to one type of process at a time, thus adding an extra effective bound.
These relate only to $C_{LedQ}$ through the following relations:
\begin{equation}
\label{eq::vectoryukawas}
\frac12\,{C_{LedQ}^{\elltau}}= x^\text{LL}_{1\,\tau}x^\text{RR}_1-x^\text{RL}_{2\,\tau}x^\text{LR}_2\,,\qquad \; \; 
\frac12\,{C_{LedQ}^{\tauh}}= x^\text{LL}_{1 }x^\text{RR}_{1\,\tau}- x^\text{RL}_{2 }x^\text{LR}_{2\,\tau}\,.
\end{equation}
Hence, we can take the two linear combinations of Yukawa pairs present in Eq.~\eqref{eq::vectoryukawas} and define new independent pairs of couplings, $x_{1,2}^{\elltau}\equiv x^\text{LL}_{1\,\tau}x^\text{RR}_1-x^\text{RL}_{2\,\tau}x^\text{LR}_2$ and $x_{1,2}^{\tauh}\equiv x^\text{LL}_{1 }x^\text{RR}_{1\,\tau}-x^\text{RL}_{2 }x^\text{LR}_{2\,\tau}$, and set bounds on them instead.
The relations among the Yukawa pairs and the WCs are then straightforward:
\begin{alignat}{3}
\label{eq:XX}
x^\text{LL}_3x^\text{LL}_{3\,\tau}&=\frac12(C_{LQ}^{(3)}-C_{LQ}^{(1)})\,,\;\; \; \; \; 
&x^\text{RL}_2x^\text{RL}_{2\,\tau}&=C_{Ld}\,,\quad
&x^\text{LR}_2x^\text{LR}_{2\,\tau}&=C_{Qe}\,,\notag\\
\tilde{x}^\text{RL}_2\tilde{x}^\text{RL}_{2\,\tau}&=C_{Lu}\,,\quad
&x^\text{LL}_1x^\text{LL}_{1\,\tau}&=-\frac12(C_{LQ}^{(1)}+3C_{LQ}^{(3)})\,,\;\; \; \; \; 
&x^\text{RR}_1x^\text{RR}_{1\,\tau}&=-C_{ed}\,,\\
\tilde{x}^\text{RR}_1\tilde{x}^\text{RR}_{1\,\tau}&=-C_{eu}\,,\quad
&x_{1,2}^{\tauh}&=\frac12\,C_{LedQ}^{\tauh}\,,\quad
&x_{1,2}^{\elltau}&=\frac12\,C_{LedQ}^{\elltau}\,.\notag
\end{alignat}
Therefore, even though we have 11 Yukawa pairs restricted by 9 effective bounds --- which implies that not all Yukawa pairs can be constrained independently from bounds on the WCs --- by introducing $x_{1,2}^{process}$, this budget is effectively changed to 9 Yukawa pairs only.
Moreover, the constraints coming from $\ell$--$\tau$ conversion are not competitive compared to those stemming from the $\tau$ decays, which implies a softer bound on $x_{1,2}^{\elltau}$, exceeding (as in the scalar LQ scenario) for $e$--$\tau$ conversion the limits suggested by perturbativity considerations; the bounds for the probed mass (with $|xx^{\prime}| \approx 1$) indicate again that the experimental constraints expected at present from the $\ell$--$\tau$ conversion in nuclei are rather weak.
\par 
The numerical results are given in Table~\ref{tab:Belle_BelleIIVector} and follow the same pattern as in the scalar scenario.
\begin{table}[tb]
\capstart
\begin{center}
\renewcommand{\arraystretch}{1.5}
\begin{tabularx}{0.98\textwidth}
{%
    >{\centering\arraybackslash}c %
  | >{\centering\arraybackslash}X %
    >{\centering\arraybackslash}X %
  | >{\centering\arraybackslash}X %
   >{\centering\arraybackslash}c %
}
\toprule
$\tau$ decays &
\multicolumn{2}{c|}{Upper bounds on $\vphantom{\Bigg(}\displaystyle\frac{|xx^\prime|}{M_\text{V}^2}\,[10^{-3}\,\text{TeV}^{-2}]$} &
\multicolumn{2}{c}{Lower bounds on $M_\text{V}$\,[TeV]}\\
\hline
Yukawa pair & Belle & Belle II & Belle & Belle II\\
\midrule
$|x_3^\text{LL}x_{3\,\tau}^\text{LL}|$ & $15$ & $1.7$ & $8.2$ & $25$\\
$|x_2^\text{RL}x_{2\,\tau}^\text{RL}|,\,|x_1^\text{RR}x_{1\,\tau}^\text{RR}|$ & $10$ & $1.5$ & $10$ & $26$\\
$|x_2^\text{LR}x_{2\,\tau}^\text{LR}|$ & $8.3$ & $1.3$ & $11$ & $28$ \\
$|\tilde{x}_2^\text{RL}\tilde{x}_{2\,\tau}^\text{RL}|$ & $24$ & $2.5$ & $6.5$ & $20$\\
$|x_1^\text{LL}x_{1\,\tau}^\text{LL}|$ & $22$ & $3.1$ & $6.7$ & $18$\\
$|\tilde{x}_1^\text{RR}\tilde{x}_{1\,\tau}^\text{RR}|$ & $17$ & $2.1$ & $7.7$ & $22$\\
$|x_{1,2}^{\tauh}|$ & $3.1$ & $0.42$ & $18$ & $49$\\
\midrule
\midrule
$\ell$--$\tau$ conversion &
\multicolumn{2}{c|}{Upper bounds on $\vphantom{\Bigg(}\displaystyle\frac{|xx^\prime|}{M_\text{V}^2}\,[10^{0}\,\text{TeV}^{-2}]$} &
\multicolumn{2}{c}{Lower bounds on $M_\text{V}$\,[TeV]}\\
\hline
 Yukawa pair & $e$--$\tau$ & $\mu$--$\tau$ & $e$--$\tau$ & $\mu$--$\tau$\\
\midrule
$|x_{1,2}^{\elltau}|$ & $330$ & $1.5$ & $0.055$ & $0.83$\\
\bottomrule
\end{tabularx}
\end{center}
\vspace*{-0.5cm}
\caption{\label{tab:Belle_BelleIIVector}
Obtained bounds for the vector leptoquark case from the bounds determined in Ref.~\cite{Husek:2020fru}.
In the left-hand part of the table, we present {\em upper} bounds on the ratio $|xx^\prime|/M_\text{V}^2$; these numbers also correspond to {\em upper} bounds on the Yukawa pairs $|xx^{\prime}|$, assuming $M_\text{V}=1$\,TeV.
On the right, there are {\em lower} bounds on the probed energy scale of the scalar leptoquarks mediating CLFV phenomena ($\Lambda_{\text{CLFV}}=M_\text{V}$), considering $|xx^{\prime}|\approx1$.
Again, the strongest bounds found are shown, stemming mostly from the $\tau$ decay constraints, except for the last row related solely to $\ell$--$\tau$ conversion.
The values are given at the 99.8\,\% confidence level.
}
\end{table}
Note that the main differences arise for the $x^\text{LL}_3x^\text{LL}_{3\,\tau}$, $x^\text{RL}_2x^\text{RL}_{2\,\tau}$ and $x_1^\text{RR}x_{1\,\tau}^\text{RR}$ pairs, which are constrained stronger than their scalar analogues.
The combination of Yukawa pairs $x_{1,2}^{\tauh}$ is also strongly constrained.
Alternatively, we can try to get information on single Yukawa couplings.
Under the flavour structure established in Eq.~\eqref{eq::YukawaStructure}, we have 14 couplings in total.
The 4 wearing a tilde contribute in pairs in a unique way (i.e.\ not to other WCs or in combination with other Yukawas) to $C_{Lu}$ and $C_{e u}$ (see Eq.~\eqref{eq:XX}), and thus the resulting relations among the single Yukawa couplings and Wilson coefficients necessarily depend on other Yukawas.
However, the 10 remaining Yukawas contribute in different ways to the rest of Wilson coefficients: There are in total 7 effective WCs once the bounds from $\ell$--$\tau$ conversion and $\tau$ decays are distinguished.
Thus, one can express 7 out of these 10 single Yukawas in terms of WCs and 3 remaining unconstrained (free) Yukawas.
The corresponding relations are given in Appendix~\ref{s:SingleYukawas}, with $\tilde{x}_2^\text{RL}$ and $\tilde{x}_1^\text{RR}$ in the former case and $x_3^\text{LL}$, $x_1^\text{LL}$ and $x_2^\text{LR}$ in the latter chosen to be free.
In light of possible non-zero values of these WCs, the (usually stronger) bounds on the $e$- and $\mu$-involved Yukawas (chosen to be free in this setting) can be used to constrain the remaining 7.

Finally, limits on single Yukawas can also be obtained with the help of the perturbativity bounds.
Taking the upper bounds on $|yy^\prime|$ obtained above, it can be assumed that the bound on the absolute value of the target Yukawa (say $y$) can be obtained by assuming that $|y^\prime|$ equals the value given by the perturbative limit.

\section{Conclusions}
\label{sec:Conclusions}

Leptoquarks are omnipresent in the recent literature that brings up extensions of the Standard Model of particle physics.
Although their concept comes from the period where the construction of Grand Unified Theories was at its high spot~\cite{Pati:1973uk,Pati:1974yy}, they have been reborn in the last ten years as a possible explanation to the LHCb \cite{London:2021lfn} and muon $(g-2)$ anomalies~\cite{Aoyama:2020ynm,Bennett:2006fi,Abi:2021gix}.
To play this role they should have a mass of few to tens TeVs.
At present, there is no experimental evidence of their existence and present bounds by LHC indicate
$M_\text{LQ} \gtrsim 1 \, \text{TeV}$.
\par 
Learning that nature allows for neutrino mixing, the immediate rationale suggests that lepton flavour violation should be also present in processes with charged leptons.
In Ref.~\cite{Husek:2020fru}, we carried out a phenomenological model-independent analysis of charged-lepton-flavour-violating processes involving the tau lepton, namely its hadronic decays and $\ell$--$\tau$ conversion in the presence of nuclei ($\ell=e,\mu$), and established bounds on the Wilson coefficients of the corresponding $D=6$ SMEFT operators.
\par 
In this article, we have merged both lines from above.
We have considered that leptoquarks (both scalar and vectors) should be driving the dynamics of charged-lepton-flavour-violating processes and, in addition, we consider an extra input:
There is a breaking of universality related to the third lepton family with respect to the two lighter ones.
Upon the integration of the heavy LQs, we get most of CLFV $D=6$ SMEFT operators and, accordingly, we can relate the couplings of LQs to fermions with the Wilson coefficients of the SMEFT.
We can then combine these relations with our results from Ref.~\cite{Husek:2020fru}, providing relevant information (in terms of bounds) on those couplings and LQs masses.
\par 
Our main results are collected in Tables~\ref{tab:Belle_BelleIIScalar}, \ref{tab:CgammaLimitsR2_S1} and \ref{tab:Belle_BelleIIVector}.
They all show bounds on the LQ masses and the Yukawa-like couplings of LQs to SM fermions.
The first two tables correspond to scalar leptoquarks and the last one to vector leptoquarks. 
In addition, Table~\ref{tab:Belle_BelleIIScalar} shows the couplings of four-fermion operators, while Table~\ref{tab:CgammaLimitsR2_S1} shows the results related to $C_{\gamma}$.
In our analyses on the latter-mentioned WC, we have included the bounds on the processes $\tau \rightarrow \ell \gamma$ for $\ell =e,\mu$ (not considered in Ref.~\cite{Husek:2020fru}) and its leptoquark generated leading contribution that appears at the one-loop level in the perturbative expansion.
As already commented in Ref.~\cite{Husek:2020fru}, we notice the significant improvement arriving with the foreseen bounds from the expected Belle II results in the hadronic decays of the tau lepton.
\par 
Considering the Belle II prospects, our results show that the bounds on scalar leptoquarks are weaker ($M_\text{S} \gtrsim 10 \, \text{TeV}$) than those on vector leptoquarks ($M_\text{V} \gtrsim 20 \, \text{TeV}$).
The numerical values of these bounds are in the expected region where LQs could help to explain the aforementioned phenomenological anomalies.

\section*{Acknowledgements}

This work has been supported in part by
Grant No.\ MCIN/AEI/FPA2017-84445-P and by MCIN/AEI/10.13039/501100011033 Grant No.\ PID2020-114473GB-I00,\\
by PROMETEO/2017/053 and PROMETEO/2021/071 (GV),
and by the Swedish Research Council grants contract numbers 2016-05996 and 2019-03779.

\appendix

\renewcommand{\theequation}{\Alph{section}.\arabic{equation}}
\renewcommand{\thetable}{\Alph{section}.\arabic{table}}
\renewcommand{\thefigure}{\Alph{section}.\arabic{figure}}

\let\appsect\section
\renewcommand{\section}{
\setcounter{equation}{0}
\setcounter{table}{0}
\setcounter{figure}{0}
\appsect}

\section*{Appendices}

\section{Identification of the SMEFT operator basis}
\label{app:identities}

In this appendix, for completeness, we list the identities and relations used to identify the $D = 6$ four-fermion operators of the basis in Ref.~\cite{Grzadkowski:2010es} from those resulting from the integration of the leptoquark fields.
\par 
Regarding the {\em scalar} Fierz identities, we have in our case for anticommuting fields
\begin{alignat}{3}
(\bar a_\text{R}b_\text{L})(\bar c_\text{L}d_\text{R})
&=(\bar aP_\text{L}b)(\bar cP_\text{R}d)
&&=-\frac12(\bar a_\text{R}\gamma_\mu d_\text{R})(\bar c_\text{L}\gamma^\mu b_\text{L})\,,\\
(\bar a_\text{R}b_\text{L})(\bar c_\text{R}d_\text{L})
&=(\bar aP_\text{L}b)(\bar cP_\text{L}d)
&&=-\frac12\bigg[
(\bar a_\text{R}d_\text{L})(\bar c_\text{R}b_\text{L})
+\frac14(\bar a_\text{R}\sigma_{\mu\nu}d_\text{L})(\bar c_\text{R}\sigma^{\mu\nu}b_\text{L})
\bigg]\,,
\end{alignat}
with $\sigma_{\mu\nu}\equiv\frac i2[\gamma_\mu,\gamma_\nu]$ and projectors $P_\text{R,L}=\frac12(1\pm\gamma_5)$.
The {\em vector} Fierz identities then read
\begin{alignat}{3}
(\bar a_\text{L}\gamma_\mu b_\text{L})(\bar c_\text{L}\gamma^\mu d_\text{L})
&=(\bar a\gamma_\mu P_\text{L}b)(\bar c\gamma^\mu P_\text{L}d)
&&=(\bar a_\text{L}\gamma_\mu d_\text{L})(\bar c_\text{L}\gamma^\mu b_\text{L})\,,\\
(\bar a_\text{L}\gamma_\mu b_\text{L})(\bar c_\text{R}\gamma^\mu d_\text{R})
&=(\bar a\gamma_\mu P_\text{L}b)(\bar c\gamma^\mu P_\text{R}d)
&&=-2(\bar a_\text{L}d_\text{R})(\bar c_\text{R}b_\text{L})\,.
\end{alignat}

The fields in the charge-conjugation basis are defined as $\psi^\text{C}\equiv C{\overline\psi}^\mathsf{T}=C\gamma_0^\mathsf{T}\psi^*$ (and consequently $\overline{\psi^\text{C}}=-\psi^\mathsf{T}C^{-1}$), with $C$ being the charge-conjugation operator (in standard representation, $C=i\gamma_2\gamma_0$).
\par
For the left-handed ($\psi_\text{L}=P_\text{L}\psi$) and right-handed ($\psi_\text{R}=P_\text{R}\psi$) components of Dirac fields, using $C^{-1}\gamma_5 C=+\gamma_5^\mathsf{T}$, we then have%
\footnote{We write $\psi_\text{L,R}^C \equiv (\psi_\text{L,R})^C$.}
\begin{alignat}{8}
\psi_\text{L}&=P_\text{L}\psi\,,\qquad
&&\psi_\text{R}&&=P_\text{R}\psi\,,\qquad
&&\overline{\psi_\text{L}}&&=\overline\psi P_\text{R}\,,\qquad
&&\overline{\psi_\text{R}}&&=\overline\psi P_\text{L}\,,\\
\psi_\text{L}^\text{C}&=P_\text{R}\psi^\text{C}\,,\qquad
&&\psi_\text{R}^\text{C}&&=P_\text{L}\psi^\text{C}\,,\qquad
&&\overline{\psi_\text{L}^\text{C}}&&=\overline{\psi^\text{C}}P_\text{L}\,,\qquad
&&\overline{\psi_\text{R}^\text{C}}&&=\overline{\psi^\text{C}}P_\text{R}\,.
\end{alignat}
Using $C^{-1}\gamma_\mu C=-\gamma_\mu^\mathsf{T}$, for anticommutating $\psi_1$ and $\psi_2$ (and thus including an additional minus sign) we arrive at
\begin{align}
\overline{\psi_1^\text{C}}\psi_2^\text{C}
&=+\overline{\psi_2}\psi_1\,,\\
\overline{\psi_1^\text{C}}\gamma_\mu\psi_2^\text{C}
&=-\overline{\psi_2}\gamma_\mu\psi_1\,,\\
\overline{\psi_1^\text{C}}\sigma_{\mu\nu}\psi_2^\text{C}
&=-\overline{\psi_2}\sigma_{\mu\nu}\psi_1\,.
\end{align}
With any representation in which $C^\dag=C^{-1}$, we have employed $\Big(\overline{\psi_1^\text{C}}\,\Gamma\,\psi_2\Big)^\dag=\overline{\psi_2}\,\Gamma\,\psi_1^\text{C}$, with $\Gamma\in\{\mathbb{1},\gamma_\mu,\sigma_{\mu\nu}\}$.
\par 
For the SU(2) indices we make use of the following manipulations:
\begin{equation}
\sum_{k=1}^3(\varepsilon\cdot\tau_k)_{ab}(\varepsilon\cdot\tau_k)_{cd}^\dag
=\sum_{k=0,1,3}\tau_{k,ab}\tau_{k,cd}
=2\delta_{ad}\delta_{bc}-\tau_{2,ab}\tau_{2,cd}
=2\delta_{ad}\delta_{bc}+\varepsilon_{ab}\varepsilon_{cd}\,,\\
\end{equation}
where $\varepsilon = i \tau_2$, and we have used the completeness relation $\sum_{k=0}^3\tau_{k,ab}\tau_{k,cd}=2\delta_{ad}\delta_{bc}$.
Combining $\varepsilon_{ab}\varepsilon_{cd}=\delta_{ac}\delta_{bd}-\delta_{ad}\delta_{bc}$ and $\sum_{k=1}^3\tau_{k,cb}\tau_{k,da}=2\delta_{ca}\delta_{bd}-\delta_{cb}\delta_{da}$, it is apparent that
\begin{equation}
\varepsilon_{ab}\varepsilon_{cd}
=\frac12\left[\left(\sum_{k=1}^3\tau_{k,cb}\tau_{k,da}\right)-\delta_{ad}\delta_{bc}\right],
\label{eq:epseps}
\end{equation}
so we find
\begin{equation}
\sum_{k=1}^3(\varepsilon\cdot\tau_k)_{ab}(\varepsilon\cdot\tau_k)_{cd}^\dag
=\delta_{ac}\delta_{bd}+\delta_{ad}\delta_{bc}
=\frac32\,\delta_{ad}\delta_{bc}+\frac12\sum_{k=1}^3\tau_{k,cb}\tau_{k,da}\,.
\label{eq:etet}
\end{equation}
Eqs.~\eqref{eq:epseps} and \eqref{eq:etet} get handy when rewriting the expressions in terms of the operators from the SMEFT basis~\cite{Grzadkowski:2010es} we work with.

\section{Leading contribution to \texorpdfstring{\boldmath $C_{\gamma}$}{Cgamma} from leptoquarks}
\label{app:CgammaMatching}

We will study the process $\ell_1 \rightarrow \ell_2 \gamma$ at the leading one-loop order and driven by scalar leptoquarks 
(see Table~\ref{tab:LQ}).
Then, by integrating out the LQs we will match our UV theory with the corresponding operators in 
the SMEFT \cite{Grzadkowski:2010es}.
This procedure will allow us to identify the leptoquark Lagrangian parameters with the $C_{\gamma}$
Wilson coefficient (defined by Eq.~(\ref{eq:RedefinitionDipoleWC})).
Once the matching is performed, we will be able to translate the bounds on the latter WC obtained in Ref.~\cite{Husek:2020fru} over the relevant leptoquark parameters.
The SMEFT operator, after spontaneous symmetry breaking, reads
\begin{equation}
\mathcal{L}\supset \frac{C_{\gamma} v}{\sqrt{2} \Lambda^{2}_\text{CLFV}}\Big[ \bar{\ell}_{2} \sigma^{\mu \nu}P_{R} \ell_{1}+\bar{\ell}_{2} \sigma^{\mu \nu}P_{L} \ell_{1} \Big]F_{\mu \nu}\, ,
\end{equation}
where $\Lambda_\text{CLFV}$ stands for the new-physics energy scale where the CLFV phenomena would take place and $F_{\mu \nu}$ is the photon field-strength tensor.
\par
\begin{figure}[t]
    \centering
    \begin{subfigure}[t]{0.45\columnwidth}
    \centering
    \includegraphics[width=0.8\columnwidth]{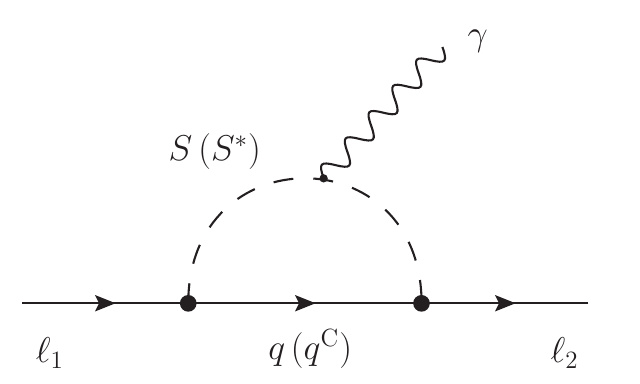}
    \caption{}
    \label{fig:llg_a}
    \end{subfigure}
    \begin{subfigure}[t]{0.5\columnwidth}
    \centering
    \includegraphics[width=0.8\columnwidth]{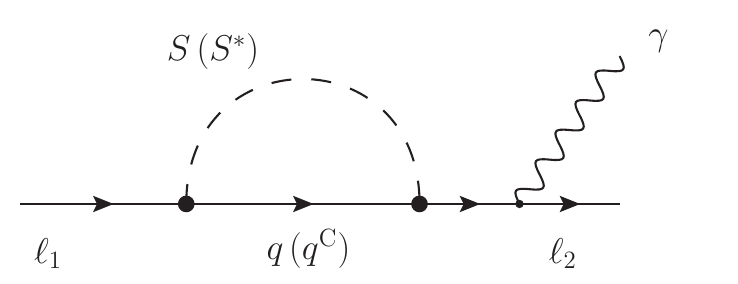}
    \caption{}
    \label{fig:llg_b}
    \end{subfigure}
    
    \begin{subfigure}[t]{0.45\columnwidth}
    \centering
    \includegraphics[width=0.8\columnwidth]{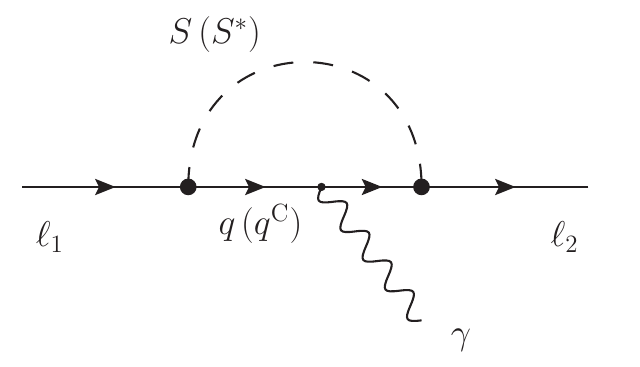}
    \caption{}
    \label{fig:llg_c}
    \end{subfigure}
    \begin{subfigure}[t]{0.5\columnwidth}
    \centering
    \includegraphics[width=0.8\columnwidth]{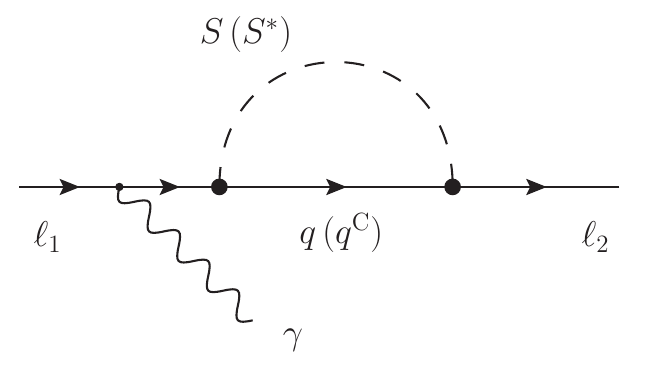}
    \caption{}
    \label{fig:llg_d}
    \end{subfigure}
\caption{Feynman diagrams of the leading-order contribution to the $\ell_{1} \to \ell_{2} \gamma$ process from scalar leptoquarks.}
\label{fig:llg}
\end{figure}

 
We write the effective Lagrangian generated by scalar leptoquarks giving the $\ell_{1} \to \ell_{2} \gamma$ process as
\begin{equation}
\mathcal{L}_\text{eff}^{\ell \to \ell^{\prime} \gamma }
=\frac{e}{2}\,\bar{\ell}_{2}\,i\sigma^{\mu \nu}\Big( \sigma_\text{R}^{\ell_{1} \ell_{2}} P_\text{R}+\sigma_\text{L}^{\ell_{1} \ell_{2}} P_\text{L} \Big) \ell_{1} F_{\mu \nu}\, ,
\end{equation}
where $\sigma_\text{R}^{\ell_{1} \ell_{2}}$ and $\sigma_\text{L}^{\ell_{1} \ell_{2}}$ are different loop functions given, in the most general case, by Ref.~\cite{Lavoura:2003xp} and recast here below in terms of the parameters of our LQ framework.
Note that in our previous work~\cite{Husek:2020fru} no distinction between left and right polarisations was considered, which in turn meant working with symmetric WC flavour matrices, e.g.\ $C_{\gamma}^{\mu \tau}=C_{\gamma}^{\tau \mu}$ (omitting quark-flavour indices).
However, within the leptoquark framework this is not always the case as it can be seen in the distinction between $\ell$--$\tau$ conversion and hadronic $\tau$ decay processes for some of the Yukawa pairs (see Sections~\ref{sec:ScalarLeptoquarks} and~\ref{sec:VectorLeptoquarks}).
This also happens in this loop computation entailing a distinction between $C_{\gamma}^{\tauh}$ and $C_{\gamma}^{\elltau}$.
Therefore, we made use of the amplitude squared of the $\ell_{1} \rightarrow \ell_{2} \gamma$ process to relate both frameworks since no direct matching at the Lagrangian level is possible.
We find the relation
\begin{equation}
\label{eq::CgammaMatching}
\bigg(\frac{C_{\gamma}^{\ell_{1} \ell_{2}}}{\Lambda_{\text{CLFV}}^{2}}\bigg)^{2}
=\frac{e^2}{\sqrt{2} v^2}\bigg[\bigg(\sum_{i}\sigma_{\text{L} , i}^{\ell_{1} \ell_{2}}\bigg)^{2}+\bigg(\sum_{i}\sigma_{\text{R} , i}^{\ell_{1} \ell_{2}}\bigg)^{2}\bigg]\, ,
\end{equation}
with $i$ running over all contributing leptoquarks, i.e.\ $S_3^{1/3}$, $R_{2}^{5/3}$ and $S_{1}^{1/3}$.
The superscripts on the WC $C_\gamma$ label the specific process for which the matching is computed --- either $\ell_{1}=\ell$, $\ell_{2}=\tau$ for $\ell$--$\tau$ conversion in nuclei, or $\ell_{1}=\tau$, $\ell_{2}=\ell$ for the hadronic $\tau$ decays, with $\ell=e,\mu$.
As explained in the main text, the $S_{3}^{1/3}$ leptoquark does not provide a chirality enhancement effect and so its $\sigma$ loop functions depend only on the lepton masses and not on the top-quark mass.
From now on, we will consider the limit of massless leptons and neglect thus the $S_{3}^{1/3}$ contribution.
Upon the integration of the leptoquarks present in the diagrams in Fig.~\ref{fig:llg}
--- following the so-called ``integration by regions'' method~\cite{Fuentes-Martin:2016uol} --- we find for the $\sigma$ loop functions%
\footnote{Note that only diagrams in Figs.~\ref{fig:llg_a} and \ref{fig:llg_c} contribute to the matching, while those in Figs.~\ref{fig:llg_b} and \ref{fig:llg_d} are responsible for the cancellation of the divergences.}
\begin{itemize}
 \item $R_{2}^{5/3}$
\begin{equation}
 \begin{split}
\sigma_{\text{L},\,R_{2}^{5/3}}^{\tauh}
=\sigma_{\text{R},\,R_{2}^{5/3}}^{\elltau}
=\frac{i N_C}{16 \pi^{2}}\frac{m_{t}}{M_\text{S}^{2}}y_{2 \, \tau}^\text{RL}y_{2}^\text{LR}V_{tb}
\bigg(\frac{3}{2}Q_{t}-\frac{1}{2}Q_{\hspace{-.25ex}R_{2}^{5/3}}\bigg) \, ,\\
\sigma_{\text{R},\,R_{2}^{5/3}}^{\tauh}
=\sigma_{\text{L},\,R_{2}^{5/3}}^{\elltau}
=\frac{i  N_C}{16 \pi^{2}}\frac{m_{t}}{M_\text{S}^{2}}y_{2\,\tau}^\text{LR}y_{2}^\text{RL}V_{tb}
\bigg(\frac{3}{2}Q_{t}-\frac{1}{2}Q_{\hspace{-.25ex}R_{2}^{5/3}}\bigg)\, ,
\end{split}
\end{equation}
\item $S_{1}^{1/3}$
\begin{equation}
 \begin{split}
\sigma_{\text{L},\,S_{1}^{1/3}}^{\tauh}
=\sigma_{\text{R},\,S_{1}^{1/3}}^{\elltau}
=-\frac{i N_C}{16 \pi^{2}}\frac{m_{t}}{M_\text{S}^{2}}y_{1 \, \tau}^\text{LL}y_{1}^\text{RR}V_{tb}
\bigg(\frac{3}{2}Q_{\bar{t}}-Q_{\hspace{-.25ex}S_{1}^{1/3}}\bigg) \, ,\\
\sigma_{\text{R},\,S_{1}^{1/3}}^{\tauh}
=\sigma_{\text{L},\,S_{1}^{1/3}}^{\elltau}
=-\frac{i N_C}{16 \pi^{2}}\frac{m_{t}}{M_\text{S}^{2}}y_{1 \, \tau}^\text{RR}y_{1}^\text{LL}V_{tb}
\bigg(\frac{3}{2}Q_{\bar{t}}-Q_{\hspace{-.25ex}S_{1}^{1/3}}\bigg)\, .
\end{split}
\end{equation}
\end{itemize}
This coincides --- in the limit of massless leptons --- with the results from Ref.~\cite{Lavoura:2003xp}, once the logarithms (stemming from the low-energy behaviour and, hence, not contributing to the matching) are removed.
Above, $m_t$ is the mass of the top quark, $V_{tb}$ is the top--bottom entry of the CKM matrix, $Q_{t}=2/3$ ($Q_{\bar{t}}=-2/3$) is the charge of the top (anti-top) quark and $Q_{\hspace{-.25ex}R_{2}^{5/3}}=5/3$ and $Q_{\hspace{-.25ex}S_{1}^{1/3}}=1/3$ are the charges of the $R_{2}^{5/3}$ and $S_{1}^{1/3}$ leptoquarks, respectively.
Note that $\sigma_{\text{L},i}^{\tauh}=\sigma_{\text{R},i}^{\elltau}$ holds true only in the case of massless leptons.
\par 
The total result for the matching, including the contribution from both leptoquarks, is thus (see Eq.~\eqref{eq::CgammaMatching})
\begin{equation}
\label{eq::FinalCgammaMatching}
\begin{split}
&\Big(\frac{C_{\gamma}}{\Lambda_{\text{CLFV}}^{2}}\Big)^{2}
=\frac{e^{2} N_C^2 m_{t}^{2}V_{tb}^{2}}{2^{11}\pi^{4}v^{2}M_\text{S}^{4}}\\
&\times\Bigg[ \Big( (y_{2 \, \tau}^\text{RL}y_{2}^\text{LR})^{2}+(y_{2 \, \tau}^\text{LR}y_{2}^\text{RL})^{2} \Big) \Big( Q_{\hspace{-.25ex}R_{2}^{5/3}}-3 Q_{t} \Big)^{2}
+ \Big( (y_{1 \, \tau}^\text{LL}y_{1}^\text{RR})^{2}+(y_{1 \, \tau}^\text{RR}y_{1}^\text{LL})^{2} \Big) \Big( Q_{\hspace{-.25ex}S_{1}^{1/3}}-3 Q_{\bar{t}} \Big)^{2}\\
&-2\Big( (y_{2 \, \tau}^\text{RL}y_{2}^\text{LR})(y_{1 \, \tau}^\text{LL}y_{1}^\text{RR})+ (y_{2 \, \tau}^\text{LR}y_{2}^\text{RL})(y_{1 \, \tau}^\text{RR}y_{1}^\text{LL}) \Big) \Big( Q_{\hspace{-.25ex}R_{2}^{5/3}}-3 Q_{t} \Big) \Big( Q_{\hspace{-.25ex}S_{1}^{1/3}}-3 Q_{\bar{t}} \Big) \Bigg]\, .
\end{split}
\end{equation}
Above, we have omitted the superscripts $\ell_{1}\ell_{2}$ of $C_{\gamma}$ since in the limit of massless leptons we obtain the same matching for hadronic $\tau$ decays and $\ell$--$\tau$ conversion in nuclei.
Consequently, the bounds from both processes can be applied.

\section{Single Yukawas of vector leptoquarks}
\label{s:SingleYukawas}

The integration of leptoquarks and the following matching to the SMEFT lead to relations between the Wilson coefficients of this EFT and products of Yukawa leptoquark couplings.
However, due to the rich variety of contributions of the Yukawas to the WCs, and the current and expected bounds on the latter from CLFV-$\tau$ processes, one can do better than just constraining pairs of Yukawas.
Under a choice of a few free Yukawas, the rest can be related to these and the WCs.
The most general leptoquark--matter interacting model provides, for vector leptoquarks, fourteen Yukawa couplings.
These are reduced, upon LQ integration, to eight different WCs which receive a total of nine bounds from the charged-lepton-flavour-violating $\tau$ processes considered in this work.
Therefore, by choosing $\tilde{x}_2^\text{RL}$ and $\tilde{x}_1^\text{RR}$ on one hand and $x_3^\text{LL}$, $x_1^\text{LL}$ and $x_2^\text{LR}$ on the other to be free, we find the trivial relations
\begin{equation}
\tilde{x}_{2\,\tau}^\text{RL}=\frac{C_{Lu}}{\tilde{x}_2^\text{RL}}\, , \qquad \tilde{x}_{1\,\tau}^\text{RR}=-\frac{C_{eu}}{\tilde{x}_1^\text{RR}}\, ,
\end{equation}
\begin{equation}
x_{3\,\tau}^\text{LL}=\frac{C_{LQ}^{(3)}-C_{LQ}^{(1)}}{2 x_3^\text{LL}}\,,\qquad
x_{1 \,\tau}^\text{LL}=-\frac{C_{LQ}^{(1)}+3C_{LQ}^{(3)}}{2 x_1^\text{LL}}\,,\qquad
x_{2\,\tau}^\text{LR}=\frac{C_{Qe}}{x_2^\text{LR}}\,,
\end{equation}
and two different solutions for
\begin{equation}
\begin{split}
x_{2\, \pm}^\text{RL}&=\frac{2C_{ed}C_{LQ}^{(1)}-C_{LedQ}^{\elltau}C_{LedQ}^{\tauh}+6C_{ed}C_{LQ}^{(3)}-4C_{Ld}C_{Qe}\mp \sqrt{A}}{4 C_{LedQ}^{\elltau}C_{Qe}}\,x_2^\text{LR}\,,\\
x_{2\,\tau\,\pm}^\text{RL}&=\frac{2C_{ed}C_{LQ}^{(1)}-C_{LedQ}^{\elltau}C_{LedQ}^{\tauh}+6C_{ed}C_{LQ}^{(3)}-4C_{Ld}C_{Qe}\pm \sqrt{A}}{4 C_{LedQ}^{\tauh}}\frac{1}{x_2^\text{LR}}\,,\\
x_{1\,\pm}^\text{RR}&=-\frac{2C_{ed}C_{LQ}^{(1)}+C_{LedQ}^{\elltau}C_{LedQ}^{\tauh}+6C_{ed}C_{LQ}^{(3)}-4C_{Ld}C_{Qe}\pm \sqrt{A}}{2 C_{LedQ}^{\tauh}(C_{LQ}^{(1)}+3C_{LQ}^{(3)})}\,x_1^\text{LL}\,,\\
x_{1\,\tau\,\pm}^\text{RR}&=\frac{2C_{ed}C_{LQ}^{(1)}+C_{LedQ}^{\elltau}C_{LedQ}^{\tauh}+6C_{ed}C_{LQ}^{(3)}-4C_{Ld}C_{Qe}\mp \sqrt{A}}{4 C_{LedQ}^{\elltau}}\frac{1}{x_1^\text{LL}}\,,
\end{split}
\label{eq:x12pm}
\end{equation}
where either all the positive or negative solutions are to be chosen, with
\begin{equation}
A=\big[C_{LedQ}^{\elltau}C_{LedQ}^{\tauh}-2 C_{ed}(C_{LQ}^{(1)}+3C_{LQ}^{(3)})+4 C_{Ld}C_{Qe}\big]^2
-16C_{Ld}C_{LedQ}^{\elltau}C_{LedQ}^{\tauh}C_{Qe}\,.
\end{equation}
Note that (usually stronger) limits on the five free variables $\tilde{x}_2^\text{RL}$, $\tilde{x}_1^\text{RR}$, $x_3^\text{LL}$, $x_1^\text{LL}$ and $x_2^\text{LR}$ can be found elsewhere, since they involve just the $e$ and $\mu$ leptons.
These can then be used, under the assumptions taken in this work, to constrain the rest of the Yukawas.

\section{Translating the bounds}
\label{app:limits}

In this appendix, we show straightforwardly how, numerically, the WC bounds are translated into limits on Yukawa pairs according to Eqs.~\eqref{eq:YY} and \eqref{eq:XX}.
The Wilson coefficients were given in the previous work as normal (Gaussian) probability distributions with mean $\mu$ and variance $\sigma^2$, i.e.\ $C=\mathcal{N}(\mu,\sigma^2)$.
The non-vanishing correlations among these were collected in the covariance matrix $W_{i j}$.
\par 
In general, given a set of functions of the WCs for which we do not know their probability distribution functions (p.d.fs.), namely $(zz^\prime)_{k}=F_{k}(\vec{C})$, with $\vec{C}=(C_{Q_{1}}, C_{Q_{2}}, \dots)$ containing all WCs considered in this work and where the symbolic notation $zz^{\prime}$ can stand both for scalar $(y)$ and vector $(x)$ Yukawa pairs, one can approximate the expectation value of $F_{k}$ and its covariance matrix through
\begin{equation}
\label{eq:ExpectedValue}
E\big[F_{k}(\vec{C})\big]\simeq F_{k}(\vec{\mu})\, ,
\end{equation}
and
\begin{equation}
\label{eq:CovarianceMatrix}
U_{k m}\equiv \text{cov}\big[F_{k}F_{m}\big]\simeq \sum_{i, j=1}^{n}\Bigg[\frac{\partial F_{k}}{\partial C_{Q_{i}}} \frac{\partial F_{m}}{\partial C_{Q_{j}}} \Bigg]_{\vec{C}=\vec{\mu}}W_{ij}\, .
\end{equation}
Above, the derivatives should be evaluated at the mean values of the WCs collected within the vector $\vec{\mu}$, which is, in our case, just a zero-valued vector $\vec\mu=\vec 0$\,.
\par
For the simple case of Eqs.~\eqref{eq:YY} and \eqref{eq:XX}, the resulting combinations of the WCs contributing to the Yukawa pairs $zz^{\prime}=\sum_{Q}a_{Q}C_{Q}$ lead again to a Gaussian p.d.f.\ for the latter, and Eqs.~\eqref{eq:ExpectedValue} and Eqs.~\eqref{eq:CovarianceMatrix} give
\begin{equation}
zz^{\prime}
=\mathcal{N}\bigg(\sum_Qa_Q\mu_Q,\,
\sum_Qa_Q^2\sigma_Q^2\bigg)
\end{equation}
for non-correlated WC, and
\begin{equation}
zz^{\prime}
=\mathcal{N}\bigg(\sum_Qa_Q\mu_Q,\,
\sum_Qa_Q^2\sigma_Q^2+2\mathop{\sum}_{Q_1<Q_2} a_{Q_1}\sigma_{Q_1}\rho_{Q_1Q_2}a_{Q_2}\sigma_{Q_2}\bigg)
\end{equation}
for the correlated ones, with $\rho_{Q_1Q_2}$ being the WC correlation matrix.


\renewcommand{\raggedright}{}

\providecommand{\href}[2]{#2}\begingroup\raggedright\endgroup

\end{document}